\documentclass[10pt, draft]{article}
\usepackage{amsfonts}
\usepackage{amssymb}
\usepackage{amsthm}
\usepackage{amsmath} 
\usepackage{graphicx}
\usepackage{color}
\date{}
\newcommand{\sss}{\setcounter{equation}{0}}
\newtheorem{theorem}{THEOREM}[section]

\newtheorem{lemma}[theorem]{LEMMA}

\newtheorem{remark}[theorem]{REMARK}

\newtheorem{prop}[theorem]{PROPOSITION}

\def\beq{\begin{equation}}
\def\ene{\end{equation}}

\newcommand{\bull}{\hfill $\Box$}

\begin{document}
\baselineskip=20 pt
\parskip 6 pt

\title{Scattering Theory for the  matrix Schr\"odinger operator on the half line with general boundary conditions
\thanks{ PACS classification (2010) 02.30.Zz; 03.65.-w; 03.65.Ge; 03.65.Nk. Mathematics Subject Classification (2010): 34L25; 34L40; 81U05; 81Uxx. Research partially supported by project  PAPIIT-DGAPA UNAM  IN102215}}
\author{ Ricardo Weder\thanks {Fellow, Sistema Nacional de Investigadores.}\thanks{ Electronic mail: weder@unam.mx. Home page: http://www.iimas.unam.mx/rweder/rweder.html} \\
Departamento de F\'{\i}sica Matem\'atica.\\
 Instituto de Investigaciones en Matem\'aticas Aplicadas y en
 Sistemas. \\
 Universidad Nacional Aut\'onoma de M\'exico.\\
  Apartado Postal 20-126,
M\'exico DF 01000, M\'exico.}

\maketitle

\vspace{.5cm}
 \centerline{{\bf Abstract}}
  We study the stationary scattering theory  for the matrix Schr\"odinger equation on the half line, with the most general boundary condition at the origin, and with  integrable selfadjoint matrix  potentials. We prove the  limiting absorption principle, we  construct the generalized Fourier maps, and we prove that they are partially isometric with initial space the subspace of absolute continuity of the matrix Schr\"odinger operator and final space $L^2((0, \infty))$. We prove the existence and the completeness of the wave operators and we establish that they are given by the stationary formulae.  We also construct the spectral shift function and we give its high-energy asymptotics. Furthermore, assuming that the potential also has a finite first moment, we prove a Levinson's theorem for the spectral shift function.

\bigskip

\section{Introduction}\sss
Let us consider the matrix Schr\"odinger operator on the half line
\begin{equation}
H_{A,B}\psi:= -\psi''+V(x)\,\psi,\qquad x\in(0,\infty).\label{1.1}
\end{equation}
The prime denotes the derivative with respect to
the spatial coordinate $x$. The wavefunction $\psi(x)$ 
will be either an $n\times n$ matrix-valued function
or it will be a column vector with $n$
components.  It is known that the most general selfadjoint boundary condition at $x=0$ for the operator \eqref{1.1} can be formulated in several equivalent way, see \cite{11}-\cite{23}.  However, It was proved in \cite{19}-\cite{23} that, without losing generality, it is useful  to state them  in terms of constant $n\times n$ matrices $A$ and $B$ as follows, 
\begin{equation}-B^\dagger\psi(0)+A^\dagger\psi'(0)=0,\label{1.2}
\end{equation}
\begin{equation}-B^\dagger A+A^\dagger B=0,\label{1.3}\end{equation}
\begin{equation}A^\dagger A+B^\dagger B>0.\label{1.4}\end{equation}
Note that  $A^\dagger B$ is selfadjoint and
the selfadjoint matrix $(A^\dagger
A+B^\dagger B)$ is positive.

The  matrices $A, B$ are not uniquely defined. We can multiply them on the right by an invertible matrix $T$ without affecting  \eqref{1.2}, \eqref{1.3} and \eqref{1.4}, and furthermore,
\beq\label{1.5}
H_{A\,T, B\,T}= H_{A,B}.
\ene

The potential $V(x)$ is a $n\times n$ selfadjoint matrix-valued function that is integrable in $(0, \infty)$, that is to say,  each entry of the matrix $V$ is Lebesgue measurable on
$(0, \infty)$ and
\begin{equation}\int_0^\infty
dx\,||V(x)||<+\infty.\label{1.6}\end{equation} 
Here,  $||V(x)||$ designates  the norm of  $V(x)$ as an operator on $\mathbf C^n$.
Of course, a matrix-valued function is integrable in $(0, \infty)$
if and only if each entry of that matrix is integrable on $(0, \infty)$.

We suppose that  $V$ is  selfadjoint,

\begin{equation}V(x)=V(x)^\dagger,\qquad x\in{\bf R}^+.\label{1.7}\end{equation}
By  the dagger we designate the matrix adjoint .

There is currently a considerable interest on this problem.  Matrix Schr\"odinger operators on the half line are important in quantum mechanical scattering of particles with internal structure, in quantum graphs and in quantum wires. See, for example, \cite{11}- \cite{10} and \cite{5}- \cite{18},  and the references quoted there.The matrix Schr\"odinger operator  on the half line \eqref{1.1} corresponds to   a star graph. It describes the behavior of $n$ connected very thin quantum wires that form a star-graph,   i.e. a graph with only one vertex and a finite number of edges of infinite length. The boundary conditions \eqref{1.2}, \eqref{1.3} and \eqref{1.4}
impose  restrictions on the value of the wave function and of its derivatives at the vertex.
The problem has physical relevance to designing elementary gates in quantum computing and nanotubes for microscopic
electronic devices, where, for example, strings of atoms may form a star-shaped graph.   
The consideration of general boundary conditions at the vertex, rather than, say, just Dirichlet boundary condition is relevant. For quantum graphs are important  boundary conditions at the vertices that link the values, at the different edges, of the wave function and of the first derivatives. An important example is the Kirchoff boundary condition that amounts to the continuity of the wave function and the conservation of the current at the vertex. Actually, a quantum graph is an idealization of wires with a small cross section that meet at vertices. The quantum graph is obtained in the limit when the cross section goes to zero. The boundary condition on the graph's vertices depends in how the limit is taken. From this point of view, it is of interest to study all the selfadjoint boundary conditions \eqref{1.2}, ]\eqref{1.3}, \eqref{1.4} as they can appears in various limit procedures.

in \cite{19} the matrix Schr\"odinger equation in the half line was considered with a self-adjoint matrix potential and with the most general self-adjoint boundary condition at the origin. For  matrix  potentials that are  integrable and have a finite first moment, it was  proven  that the  scattering matrix is continuous at zero energy, and  an explicit formula for its value at zero energy was given. Also, the  small-energy asymptotics was   established  for the  Jost matrix, its inverse, and  other quantities that are important in  direct and inverse scattering problems. The  paper \cite{19}  is complementary  to the classical monograph  by Agranovich and Marchenko \cite{20} that only considers the  Dirichlet boundary condition, but where the behavior  at  zero energy is studied. The article \cite{19} is also  complementary  to the results by Harmer, \cite{8,9,10} where matrix Schr\"odinger operators on the half line with general boundary conditions are investigated,  but the small-energy analysis is not considered. 

The paper  \cite{21} considers the high-energy asymptotics  for the Jost matrix, for its  inverse and for the scattering matrix.  Also   Levinson's theorem  was proven. This theorem relates the number of bound states to the change in the argument of the determinant of the scattering matrix. The article \cite{21} complements  Agranovich and Marchenko's results \cite{20} who gave the large energy asymptotics and Levinson's theorem, but  only under the Dirichlet boundary condition. The contribution \cite{21}  also complements the study \cite{8,9,10} by Harmer (see also \cite{22}), where the general selfadjoint boundary condition  is considered but the large-energy  asymptotics of the scattering matrix is obtained by only providing the leading term, with the remaining terms given  as $o(1)$ for large energy. In \cite{22},  the leading term is given, but we also obtain  the next-order term, that behaves as $ 1/ k$ with  $k$ the square root of the energy  and  the large-energy  asymptotics up to $O(1/k^2)$ is obtained,  which is crucial in establishing the Fourier transform of several  quantities that are important in the  inverse scattering problem. In \cite{23} the results of \cite{19} are improved in the case where the potential is integrable and has a finite second moment. It is proven 
that the  scattering matrix is differentiable at zero energy, and an  explicit
formula is given  for its derivative at zero energy. Also, \cite{23} improves, in the case where the second moment of the potential is finite,  the small -energy asymptotics given in \cite{19} for the Jost matrix, its inverse, and  other quantities important in the   direct and inverse scattering problems.

In this paper we complement the results of \cite{19}-\cite{23} by studying the stationary scattering theory of the matrix Schr\"odinger operator $H_{A,B}$ in equation \eqref{1.1}, with the most general boundary condition \eqref{1.2}-\eqref{1.4} and with matrix potentials that satisfy \eqref{1.6} and \eqref{1.7}.  

The paper is organized as follows.
In Section~2 we introduce notations and definitions that we use  throughout the paper. In Section~3 we state results from \cite{19}-\cite{23}, that we need,  in the Jost solution, the regular solution, the Jost and scattering matrices and in transformations of the matrices $A,B$ that give the boundary conditions. In Section ~4 we construct a selfadjoint realization of the matrix Schr\"odinger operator \eqref{1.1}, that we also denote by  $H_{A,B}$, using quadratic forms. In Section ~5 we study the resolvent of $H_{A,B}$ and we prove the limiting absorption principle. It is customary in scattering theory to formulate the limiting absorption principle in terms of weighted $L^2$ spaces or  of a Besov space and its dual. See for example \cite{ho}-\cite{ya2}. On the contrary, since we work under the condition that the potential matrix $V$ is only integrable, we have found it more convenient to formulate the limiting absorption principle between $L^1$ and $L^\infty$.
In Section ~6 we construct the generalized Fourier maps. In Section ~7 we study the wave operators. Section ~8 is dedicated to the  scattering operator and the scattering matrix. Finally, in Section ~9 the spectral shift function is constructed and  a Levinson's theorem for the spectral shift function is proven.   

\section{Notations and definitions}\sss

We designate  by  ${\bf C}^+$ the upper-half complex plane,  by ${\bf R}$  the real axis, and we let ${\overline{{\bf C}^+}}:={\bf C}^+\cup{\bf R}$.  For any $k \in  \overline{{\bf C}^+}$ we denote by $k^*$ its complex conjugate. For any matrix $D$ we designate by $D^\dagger$ its adjoint. We denote by $C$ a positive constant that  might not take the same value when it appears in different places.

We denote by $L^p,\, 1 \leq p \leq \infty$, the standard space $L^p((0, \infty))$ of measurable functions with values in $\mathbf C^n$, in particular, we will use $L^2$ instead of $L^2((0, \infty))$, for simplicity.  

By $H_l, l=1,2, $ we denote  the Sobolev space of order $l$ of all square integrable, complex valued functions   with all distributional derivatives up to order $l$ given by square integrable functions \cite{24}. We designate by $H_{1,0}$ the completion of $C^\infty_0((0, \infty))$ in the norm of  $H_1$, where $C^\infty_0((0, \infty))$ is the space of all infinitely differentiable, complex valued functions with compact support. We use the notation,
\beq \label{2.1}
\mathbf{H}_l:=\oplus_{j=1}^n\, H_{l}, \quad l=1,2,
\ene
for the first and second Sobolev spaces of functions with values in $\mathbf C^n$.

Let $\mathbf {W}_{l,\infty}, l=1,2$ be the Sobolev space of all  measurable functions  on $(0,\infty)$  with values in $\mathbf C^n$ that together with all its derivatives of order up to $l$ are essentially bounded \cite{24}. 

 For any pair of Banach spaces $X,Y$ we designate by $\mathcal B(X,Y)$ the Banach space of all bounded linear operators from $X$ into $Y$. In the case where $X=Y$ we use $\mathcal B(X)$ instead of $\mathcal B(X,X).$

For any  Hilbert-Schmidt operator $G$ we denote by $\left\| G \right\|_{2}$  its Hilbert-Schmidt norm and for any trace class operator $G$  we denote by $\left\| G \right\|_1$ its trace norm and by $\textrm{Tr}(G)$ its trace. For any densely defined operator $D$ in a Banach space we denote by $\rho(D)$ its resolvent set, i.e., the open set of all $ z \in \mathbf C$ such that $D-z$ is invertible  and $\left( D-z\right)^{-1}$ is bounded.

For any selfadjoint operator $\mathbf H$ we denote by $\mathcal H_{\textrm ac}(\mathbf H )$ its  subspace of absolute continuity and by $P_{\textrm ac}(\mathbf{H})$ the  projector onto   $\mathcal H_{\textrm ac}(\mathbf H )$.

For any pair, $ \mathbf{H}_0, \mathbf{H}_1$  of selfadjoint operators in the same Hilbert space  we denote the wave operators by,

$$
W_\pm\left( \mathbf{H}_1, \mathbf{H}_0 \right):= \textrm{s}- \lim_{t \rightarrow \pm \infty}\, e^{it\, \mathbf{H}_1}\,  e^{-it\, \mathbf{H}_0}\, P_{\textrm ac}(\mathbf{H}_0),
$$ 
provided that the strong limits exist. The wave operators are complete if $\textrm{Range}\left[ W_\pm(\mathbf{H}_1, \mathbf{H}_0)\right]= \mathcal{H}_{\textrm ac}\left( \mathbf{H}_1 \right)$. The scattering operator is defined as,

$$
S\left( \mathbf{H}_1,  \mathbf{H}_0 \right):=\left( W_+\left( \mathbf{H}_1, \mathbf{H}_0 \right)\right)^\star\,W_-\left( \mathbf{H}_1, \mathbf{H}_0 \right).
$$

\section{Preliminary results}\sss

In this section we introduce certain results that we need. See \cite{19,21,23,20}. We always assume that  $V$ satisfies \eqref{1.6}, \eqref{1.7}. We first consider some $n\times n$ matrix solutions to the equation  
 \begin{equation}
 -\psi''+V(x)\,\psi = k^2\, \psi,\qquad x\in(0,\infty), k \in \overline{{\bf C}^+}.\label{3.1}
\end{equation}
For any pair $F,G$ of $n\times n$ matrix valued functions  defined on $(0,\infty)$,  we denote by,

$$
[F;G]:=FG'-F'G
$$
 the Wronskian. It follows from a direct calculation
that for any two $n\times n$ solutions $\phi(k,x)$ and
$\psi(k,x)$ to \eqref{3.1}, each of the Wronskians $[\phi(k^*,x)^\dagger;\psi(k,x)]$ and $[\phi(-k^*,x)^\dagger;\psi(k,x)]$ is independent of $x.$

The Jost solution to \eqref{3.1}  is the $n\times n$ matrix solution
satisfying, for $k\in{\overline{{\bf C}^+}}\setminus\{0\},$ the asymptotics

\begin{equation}f(k,x)=e^{ikx}[I_n+o(1/x)],\quad
f'(k,x)=ik\,e^{ikx}[I_n+o(1/x)],\qquad x\to+\infty,\label{3.2}
\end{equation}
where $I_n$ denotes the $n\times n$ identity matrix. It is well  is known \cite{19,20}, that for each fixed $x,$ 
$f(k,x)$ and $f'(k,x)$ are analytic for $k\in{\bf C}^+$
and continuous for $k\in{\overline{{\bf C}^+}}$.
It follows from \eqref{3.2}   that for each fixed $k \in{\bf C}^+$, each of the $n$ columns of
$f(k,x)$  decays exponentially to zero as $x\to+\infty$. 
We have that,
\begin{equation}[f(\pm k,x)^\dagger;f(\pm k,x)]=
\pm 2ikI_n,\qquad k\in{\bf R},\label{3.3}
\end{equation}
\begin{equation}[f(-k^*,x)^\dagger;f(k,x)]=
0,\qquad k\in{\overline{{\bf C}^+}}.\label{3.4}
\end{equation}

The matrix Schr\"odinger equation \eqref{3.1} also has the $n\times n$ matrix solution $g(k,x)$
that satisfies , for each $k\in{\overline{{\bf C}^+}}\setminus\{0\},$ the following asymptotics (see page 38 of \cite{20})

\begin{equation} \label{3.5}
g(k,x)=e^{-ikx}[I_n+o(1/x)],\, g'(k,x)= -i k \, e^{-i k x}\, (I+ o(1/x)),  \quad x \rightarrow \infty.
\end{equation} 

It is proven in  \cite{20} that $g(k,x)$ and $g'(k,x)$ are analytic in $k\in{\bf C}^+$
and continuous in $k\in{\overline{{\bf C}^+}}\setminus\{0\}$ for each fixed $x.$ Equation \eqref{3.5} implies  that each of the $n$ columns of $g(k,x)$  grows exponentially as $x\to+\infty$ for each fixed $k\in{\bf C}^+.$

 It follows from \eqref{3.2} and \eqref{3.5} that,
 \beq \label{3.6}
 \left[f(-k^*,x)^\dagger;  g(k,x) \right]= -2ik,\, \left[g(-k^*,x)^\dagger;  f(k,x) \right]= 2ik, \qquad  \left[g(-k^\star,x)^\dagger;  g(k,x) \right]=0.
 \ene
On page  28 of \cite{20} it is also proven that  for each
$k\in{\overline{{\bf C}^+}}\setminus\{0\},$ the combined $2n$ columns of $f(k,x)$ and of $g(k,x)$
form a fundamental set of solutions to  \eqref{3.1}. Hence,
any column-vector solution $\omega(k,x)$ to  \eqref{3.1}  can be written as a linear combination of them,

\begin{equation}\omega(k,x)=f(k,x)\,\xi+g(k,x)\,\eta,\label{3.7}
\end{equation}
for some constant column vectors $\xi$ and $\eta$ in ${\bf C}^n$.

The regular solution $\varphi_{A,B}(k,x)$ is the $n\times n$ matrix solution to \eqref{3.1}
that satisfies  the initial conditions
\begin{equation}
\varphi_{A,B}(k,0)=A,\quad \varphi_{A,B}'(k,0)=B,\label{3.8}
\end{equation}
where $A$ and $B$ are the matrices appearing in \eqref{1.2}, \eqref{1.3}, \eqref{1.4}. It is known that \cite{19},  for each fixed
$x\in{\bf R}^+,$ 
 $\varphi(k,x)$ is entire in $k$
in the complex plane ${\bf C}.$
As $k^2$ appears in \eqref{3.1} and the initial values given in \eqref{3.8}
are independent of $k$,
\begin{equation}\varphi(-k,x)=\varphi(k,x),\qquad k\in{\bf C},\quad x\in (0, \infty).\label{3.9}
\end{equation}

The Jost matrix $J(k)$ is defined as follows,
\begin{equation}
J_{A,B}(k):=[f(-k^*,x)^\dagger;\varphi_{A,B}(k,x)],\qquad k\in{\overline{{\bf C}^+}}.\label{3.10}
\end{equation}
Evaluating the Wronskian at $x=0$ and using \eqref{3.8} we obtain that,

\begin{equation}
J_{A,B}(k)=f(-k^*,0)^\dagger B-f'(-k^*,0)^\dagger A,  \qquad k\in{\overline{{\bf C}^+}}.      \label{3.11}
\end{equation}
Note that  $J(k)$ is well defined for ${\overline{{\bf C}^+}}$ because $f(-k^*,0)^\dagger$
and $f'(-k^*,0)^\dagger$ are analytic in $k\in{{\bf C}^+}$ and
continuous in $k\in\overline{{\bf C}^+}.$ Furthermore \cite{19}, $J(k)$ is invertible for $ k \in \mathbf{R}\setminus 0$ and $J(k)^{-1}$ is continuous for $ k \in \mathbf{R}\setminus 0$ .

The regular solution can be expressed in terms of the Jost solution and the Jost matrix as follows (see equation (3.5) of\cite{21}), 
\begin{equation} \label{3.12}
\varphi_{A,B}(k,x)={\displaystyle}\frac{1}{2ik}\,
f(k,x)\,J_{A,B}(-k)-{\displaystyle}\frac{1}{2ik}\,f(-k,x)\,J_{A,B}(k),\qquad k\in{\bf R}\setminus\{0\}.
\end{equation}

Let us define the physical solution $ \psi_{A,B}(k,x)$ as follows,
 \beq \label{3.13}
\psi_{A, B}(k,x):= - i k \, \varphi_{A, B}(k,x)\, J^{-1}_{A, B}(k).
\ene
Note that  there is a difference by a factor of $2$ with the physical solutions defined in equation (3.7) of \cite{21}. 
The scattering matrix is defined as,
\beq\label{3.14}
S_{A,B}(k):= - J_{A,B}(-k)\, J_{A,B}(k)^{-1}, \qquad k \in \mathbf{R}\setminus 0.
\ene
Then, the scattering matrix $S_{A,B}(k)$   is unitary for $k\in{\bf R}\setminus 0$ and it satisfies,
\begin{equation}
S(-k)=S(k)^{-1}=S(k)^\dagger,\qquad k\in{\bf R}\setminus 0.\label{3.15}\end{equation}

The physical solution  can be written in terms of the Jost solution and the scattering matrix as follows (see equation (3.10) of \cite{21}),
\beq \label{3.16}
\psi_{A,B}(k,x):= \frac{1}{2}\, f(-k,x)+ \frac{1}{2}\, f(k,x)\,S_{A,B}(k),\qquad k\in{\bf R}\setminus 0.
\end{equation}
Note that as the physical solution \eqref{3.13} differs by a factor of $2$ with the physical solutions defined in equation (3.7) of \cite{21}, equation \eqref{3.16} differs by a factor of $ 1/2$ with equation (3.10) of \cite{21}. Equations \eqref{3.12} and \eqref{3.16} hold under the assumption that $V$ is integrable. The condition that $V$ also has a finite first moment is only needed in \cite{21} in order that \eqref{3.16} also holds for $k=0$.

Let $A,B$ be the matrices that appear in the definition of the boundary conditions in equations \eqref{1.2}, \eqref{1.3} and \eqref{1.4}. In  Proposition 4.1 of \cite{21} it is proven that if $A,B$ are multiplied on the right by an invertible matrix $T$,  and if $V$ is unchanged,  the regular solution, the Jost matrix and the scattering matrix change as follows,  

\begin{align}
\varphi_{AT, BT}(k,x)&= \varphi_{A,B}(k,x)\, T,  \qquad k \in   {\bf C},   \label{3.17}\\
J_{AT,BT}(k)&= J_{A,B}(k,x)\,T,   \qquad k \in    \overline{{\bf C}^+},                  \label{3.18} \\
S_{AT,BT}(k)&= S_{A,B}(k), \qquad k \in \mathbf{R}\setminus 0. \label{3.19}
\end{align} 

Furthermore,  it is also proven in Proposition 4.1 of \cite{21} (with $M, M^\dagger$ there , replaced, respectively, by $ M^\dagger, M$ ) that under the unitary transformation $V\mapsto M V M^\dagger,$  with a unitary matrix $M,$ and the combination of three consecutive
transformation $(A,B)\mapsto (M AT_1 M^\dagger T_2,M B T_1 M^\dagger T_2),$
first by a right multiplication by an invertible matrix
$T_1,$ then by the unitary transformation with $M,$
followed by a right multiplication by an invertible matrix
$T_2,$ we have  that 
\begin{align}
f_{ M V M^\dagger}(k,x)&= M f(k,x) M^\dagger,  \qquad k \in \overline{{\bf C}^+}, \label{3.20} \\
\varphi_{M\, V M^\dagger, M AT_1 M^\dagger T_2, M BT_1 M^\dagger T_2 }(k,x)  &=M \varphi_{V, A,B}(k,x)
T_1M^\dagger T_2, \qquad k \in {\bf C}, \label{3.21}\\
J_{M\, V M^\dagger, M AT_1 M^\dagger T_2, M BT_1 M^\dagger T_2}(k,x)&= M J_{V,A,B}(k,x)\, T_1 M^\dagger T_2,   \qquad k \in    \overline{{\bf C}^+},   \label{3.22}\\
   S_{M\, V M^\dagger, M AT_1 M^\dagger T_2, M B T_1 M^\dagger T_2} &= M\, S_{V, A,B}\, M^\dagger, 
 k \in   \mathbf{R}\setminus 0. \label{3.23}
   \end{align}

In \eqref{3.20}- \eqref{3.23} we made explicit, for clarity, the dependence in $V$ of the Jost solution, the regular solution, the Jost matrix and  the scattering matrix.

Furthermore, we have that, 

\beq \label{3.24}
H_{{ M V M^\dagger, M AT_1 M^\dagger T_2, M B T_1 M^\dagger T_2}}= M \, H_{V,A,B}\, M^\dagger,
\ene
where we also made explicit the dependence in $V$ of the matrix Schr\"odinger operator \eqref{1.1}.

 The transformation $V\mapsto V$ and $(A,B)\mapsto (AT,BT)$ with
an invertible matrix $T$ is just a change of
parametrization in the boundary conditions \eqref{1.2}, \eqref{1.3} and \eqref{1.4}.
On the  contrary, the unitary
transformation $V\mapsto M\, V M^\dagger$ and $(A,B)\mapsto (M A M^\dagger, M B M^\dagger)$
with a unitary matrix $M$ is a change of representation in quantum mechanical sense.

It is useful to consider the case where the matrices $A,B$ are diagonal. This is motivated by the general selfadjoint boundary condition \cite{26,25,27} in the scalar case, i.e. when  $n=1$. We denote this special pair of diagonal matrices by  $\tilde A$ and $\tilde B,$ where

\begin{equation}\tilde {A}:=-\mathrm{diag}\{\sin\theta_1,\dots,\sin\theta_n\},
\quad \tilde B:=\mathrm{diag}\{\cos\theta_1,\dots,\cos\theta_n\}.\label{3.25}
\end{equation}
In this case the boundary conditions \eqref{1.2} are,
\beq\label{3.25b}
\cos\theta_j\, \psi_j(0)+ \sin\theta_j\, \psi_j'(0)=0, \qquad j=1,2,\cdots,n.
\ene

The real parameters $\theta_j$ take values in the interval $(0,\pi].$
The the case $\theta_j=\pi/2$ corresponds to the Neumann boundary condition, case $\theta_j=\pi$ corresponds to the Dirichlet boundary condition,  and the case where $ \theta_j \neq  \pi/2, \pi$ corresponds to mixed boundary conditions. We suppose that there
are $n_{\mathrm N}$ values with $\theta_j=\pi/2$ and $n_{\mathrm D}$ values with $\theta_j=\pi,$
and hence there are $n_{\mathrm M}$ remaining values,
with $n_{\mathrm M}:=n-n_{\mathrm N}-n_{\mathrm D},$ such that the corresponding
$\theta_j$-values lie in
the interval $(0,\pi/2)$ or $(\pi/2,\pi).$
We allow for  the special cases where any of $n_{\mathrm N},$ $n_{\mathrm D},$ and $n_{\mathrm M}$ may be zero or $n.$ The subscripts N, D, and M refer, respectively, to Neumann, Dirichlet or mixed boundary  conditions.
We order  the $\theta_j$-values in \eqref{3.25} so that the first $n_{\mathrm M}$ values of $\theta_j$ correspond to the mixed boundary conditions, the next $n_{\mathrm D}$ values correspond to the Dirichlet boundary conditions, and the remaining $n_{\mathrm N}$ values correspond to the Neumann boundary conditions.

Note that $\tilde{A}, \tilde{B}$ satisfy \eqref{1.2}, \eqref{1.3}, \eqref{1.4} with $\tilde{A}, \tilde{B}$ instead of $A,B$ there.

In Proposition 4.3 of \cite{21} it is proven that for any pair of matrices $(A,B)$ that satisfy \eqref{1.2}-\eqref{1.4} there is a pair of diagonal matrices $(\tilde{A}, \tilde{B})$ as in \eqref{3.25}, a unitary matrix $M$ and a two invertible matrices $T_1,T_2$ such that,
\beq \label{3.26}
A=    M\, \tilde{A}\, T_2 \, M^\dagger\, T_1, \quad  B=    M\, \tilde{B}\, T_2 \, M^\dagger\, T_1.
\ene
Note that $ T_1,T_2$ in \eqref{3.26} correspond, respectively to  $  T_1^{-1},T_2^{-1}$ in Proposition 4.3 of \cite{21}.

  For $V\equiv 0,$ we have that  (see Section ~V of \cite{21}),
\begin{equation}J_{0,A,B}(k)=B-ikA,\quad S_{0,A,B}(k)=-(B+ikA)(B-ikA)^{-1}.\label{3.27}
\end{equation}

Furthermore,  in the  parametrization of the boundary conditions  with $\tilde A$ and $\tilde B$
these matrices are diagonal,
\begin{equation} J_{0, \tilde{A}, \tilde{B}}(k)=\tilde B-ik\tilde A=
\mathrm{diag}\{\cos \theta_1+ik\sin\theta_1 ,\dots,\cos\theta_{n_{\mathrm M}}+ik\sin\theta_{n_{\mathrm M}},-I_{n_{\mathrm D}},ikI_{n_{\mathrm N}}\},
\label{3.28}\end{equation}
\begin{equation} J_{0, \tilde{A}, \tilde{B}}(k)^{-1}=
\mathrm{diag}\left\{{\displaystyle}\frac{1}{\cos \theta_1+ik\sin\theta_1},\dots,
{\displaystyle}\frac{1}{\cos\theta_{n_{\mathrm M}}+ik\sin\theta_{n_{\mathrm M}}},-I_{n_{\mathrm D}},{\displaystyle}\frac{1}{ik}\,I_{n_{\mathrm N}}\right\},
\label{3.29}
\end{equation}
\begin{equation} S_{0, \tilde{A}, \tilde{B}}(k)=
\mathrm{diag}\left\{{\displaystyle}\frac{-\cos \theta_1+ik\sin\theta_1}{\cos \theta_1+ik\sin\theta_1},\dots,
{\displaystyle}\frac{-\cos\theta_{n_{\mathrm M}}+ik\sin\theta_{n_{\mathrm M}}}{\cos\theta_{n_{\mathrm M}}+ik\sin\theta_{n_{\mathrm M}}},
-I_{n_{\mathrm D}},I_{n_{\mathrm N}}\right\}.
\label{3.30}
\end{equation}
Moreover,
\beq \label{3.31}
\varphi_{0,\tilde{A}, \tilde{B}}= \mathrm{diag}\{ \varphi_{0,\tilde{A}, \tilde{B},1 }, \varphi_{0,\tilde{A}, \tilde{B},2}, \cdots ,\varphi_{0,\tilde{A}, \tilde{B},n } \},
\ene
where,
\beq\label{3.32}
\varphi_{0, \tilde{A}, \tilde{B},j}= \begin{cases} \displaystyle-\frac{1}{k}\sin kx,  \,\textrm{if}\, \theta_j=\pi, \\
\displaystyle-\cos kx, \, \textrm{if}\, \theta_j= \pi/2, \\
\displaystyle\frac{1}{k} \cos\theta_j\, \sin kx -\sin\theta_j\, \cos kx,  \,\textrm{if} \,\theta_j \neq \pi, \pi/2.
\end{cases}
\ene

In Theorem 7.6 of \cite{21} it proven that (note the slight change of notation with respect to Theorem 7.6 of \cite{21})
\begin{equation}S_{A,B}(k)=S_{A,B}(\infty)+{\displaystyle}\frac{G(k)}{ik}+O(1/k^2),\qquad k\to\pm\infty,
\label{3.33}\end{equation}
where 
\beq \label{3.34}
 S_{A,B} (\infty):=MZ_0 M^\dagger,
\ene
\beq\label{3.35}
G(k):=-2MZ_1 M^\dagger+Q_1 S_{A,B}(\infty)+S_{A,B}(\infty)\,Q_1+S_{A,B}(\infty)Q_2(k)S_{A,B}(\infty)+Q_2(-k),
\ene
with $M$ the unitary matrix that appears in \eqref{3.26}, 
\beq\label{3.36}
Z_0:= \textrm{diag}\{ I_{\textrm n_M}, -I_{\textrm n_D}, I_{\textrm n_N}  \}.
\ene
\beq\label{3.37}
Z_1:=\mathrm{diag}\{\cot\theta_1,\dots,\cot\theta_{n_{\mathrm M}},0_{n_{\mathrm D}},0_{n_{\mathrm N}}\}.
\ene
For the definition of $\theta_1, \cdots \theta_{n_{\mathrm M}}, n_{\mathrm D}, n_{\mathrm N}$ see \eqref{3.25}, the text below this equation and \eqref{3.26}. Furthermore,  $Q_1$ and $Q_2(k)$ are the following matrices,  
$$
Q_1:= \frac{1}{2}\, \int_0^\infty\, dy\, V(y), \qquad Q_2(k):=   \frac{1}{2}\, \int_0^\infty\, dy\, e^{2iky}\,V(y).
$$

\section{The Hamiltonian} \sss
\subsection{The case of zero potential }
We consider first the case where the potential $V$ is zero.We denote by $H_{0, A,B}$ the self- adjoint  realization of
$ - \frac{d^2}{ d x^2} $ with the boundary condition \eqref{1.2}, namely
\beq\label{10.1}
H_{0,A,B} \psi= -\frac{d^2}{d x^2} \,\psi, \qquad \psi \in D(H_{0,A,B}),
\ene
where
\beq\label{10.2}
D(H_{0,A,B}):= \left\{ \psi \in \mathbf{H}_2 : -B^\dagger\,\psi(0)+A^\dagger\,\psi'(0)=0 \right\}.
\ene

Note that $H_{0,A T,B T}= H_{0,A,B}$ for all invertible matrices $T$. Recall that in the particular case of the diagonal matrices $ \tilde{A}, \tilde{B}$ \eqref{3.25} the boundary conditions \eqref{1.2} are given by \eqref{3.25b}. These equations  can be written as,
\beq \label{10.4}
\psi'(0)= - \cot\theta_j\, \psi_j(0), \,\mathrm{if} \,\theta_j \neq \pi, \,\mathrm{and} \qquad \psi_j(0)=0, \,\mathrm{if} \,\theta_j= \pi.
\ene

Let us construct the quadratic form associated to $ H_{0, \tilde{A}, \tilde{B}}$. We denote,
\beq \label{10.5}
H_{1,j}:= H_{1,0},\, \mathrm{if}\, \theta_j= \pi, \, \mathrm{and}\, \qquad H_{1,j}:= H_1,\, \textrm{if}\, \theta_j \neq \pi.
\ene
We designate,
\beq \label{10.6}
\mathbf{H}_{1,\tilde{A}, \tilde{B}}:= \oplus_{j=1}^n\, H_{1,j}.
\ene
We define the quadratic form with domain $\mathbf{H}_{1, \tilde{A},\tilde{B}}$,

\beq\label{10.7}
h_{0, \tilde A, \tilde B}(\varphi,\psi):= \left(\varphi', \psi'\right) -\sum_{j=1}^n \widehat{\cot} \theta_j\, \varphi_j(0) \,\overline{\psi_j(0)},
\ene
where $ \widehat{\cot} \theta_j=0$ if $ \theta_j=\pi/2$, or $\theta_j= \pi$, and $ \widehat{\cot} \theta_j= \cot \theta_j$ if $\theta_j \neq \pi/2, \pi$.

Since, for any $ \varepsilon >0$ there is a constant $K_\varepsilon$ such that,
\beq\label{10.7b}
\left| \psi(0)\right| \leq \varepsilon \left\| \psi\right\|_{\mathbf{H}_1} + K_\varepsilon\, \left\| \psi\right\|_{L^2},
\ene
the symmetric form $h_{0,\tilde A,\tilde B}$ is closed and bounded below. It follows from Theorems 2.1 and 2.6 in chapter 6 of \cite{ka} that  $H_{0, \tilde{A}, \tilde{B}}$ is the selfadjoint bounded below operator associated to the quadratic form $ h_{0 \tilde A, \tilde B}$.

We define the diagonal matrix

\beq\label{10.8}
\Theta:= \mathrm{Diag}\,\{\widehat{\cot} \theta_1, \widehat{\cot} \theta_2, \cdots \widehat{\cot}\theta_n \}.
\ene
Then, by \eqref{3.24}, \eqref{3.26}, \eqref{10.7}  the quadratic form associated to $H_{0,A,B}$ is given by,

\beq\label{10.9}
h_{0,A,B}\left(\varphi,\psi\right):= \left( \varphi',\psi' \right)- \sum_{j=1}^n \left< M \Theta M^\dagger  \varphi(0), \psi(0)      \right>,
\ene 
where by $ <\cdot,\cdot>$ we denote the scalar product in $\mathbf C^n$, and the domain of $h_{0,A,B}$ is given by

\beq \label{10.10}
D\left( h_{0,A,B}\right)= \mathbf{H}_{1,A,B}\, \qquad \textrm{where}\,\,      \mathbf{H}_{1,A,B}:=  M \mathbf{H}_{1,\tilde{A}, \tilde{B}} \subset \mathbf{H}_1.
\ene

\subsection{The case of integrable potential}
Suppose that $ V$ satisfies \eqref{1.6}, \eqref{1.7}. Let us define the following quadratic form,

\beq \label{10.11}
h_{A,B}\left(\varphi,\psi\right):= h_{0,A,B}\left( \varphi,\psi\right)+ \left(V\varphi,\psi\right), \qquad D(h_{A,B})= \mathbf{H}_{1,A,B}.
\ene
By \eqref{10.7b},  for any $ \varepsilon >0$ there is a constant $K_\varepsilon$ such that,
$$
\left|\left( V \varphi, \varphi\right)\right| \leq \varepsilon \left\| \varphi \right\|^2_{\mathbf{H}_1} + K_\varepsilon\, \left\| \varphi\right\|^2_{L^2}.
$$
Hence, the symmetric form $h_{A,B}$  is closed,  and bounded below. Let us denote by 
$H_{A,B}$ the associated bounded below selfadjoint operator (see theorems 2.1 and 2.6 in chapter 6 of \cite{ka} ) . Note that

$$
D\left( H_{A,B}  \right)= \left\{ \psi \in  \mathbf{H}_{1,A,B} : -B^\dagger\psi(0)+A^\dagger\psi'(0)=0,    - \psi'' + V \psi  
\in L^2 \right\}.
$$

\section{The Resolvent}\sss
\subsection{The case of zero potential}
We first consider the case of the diagonal matrices $\tilde{A}, \tilde{B}$ given in \eqref{3.25}. Let us denote by $ R_{0, \tilde{A}, \tilde{B}}(z)$ the resolvent of $H_{0,\tilde{A}, \tilde{B}}$,

\beq \label{11.3}
 R_{0, \tilde{A}, \tilde{B}}(z):= \left( H_{0,\tilde{A}, \tilde{B}} - z  \right)^{-1}, \quad z \in \rho\left(H_{0,\tilde{A},\tilde{B}}\right).
\ene
 Let, $R_{0, \tilde{A}, \tilde{B}, }(z)(x,y)$ be the integral kernel of  $R_{0,\tilde{A}, \tilde{B}}(z)$. Then, we have that,

\beq \label{11.4}
R_{0, \tilde{A}, \tilde{B}}(z)(x,y)=\begin{cases} \varphi_{0,\tilde{A}, \tilde{B}}(x,k)\, e^{ik y} \,\left[J_{0, \tilde{A},\tilde{B}}(k)\right]^{-1} , \quad x \leq y, \\
e^{ik x}\,   \, \varphi_{0,\tilde{A}, \tilde{B}}(y,k)\,   \left[J_{0, \tilde{A}, \tilde{B}}(k)\right]^{-1}    , \quad x \geq y, 
\end{cases}
\ene
where, $k:= \sqrt{z}, \textrm{Im}\, k \geq 0$, and $J_{0,\tilde{A}, \tilde{B}}$ is the Jost matrix given in \eqref{3.28}. 

We designate by $ R_{0, A, B}(z)$ the resolvent of $H_{0,A,B}$. Then, by \eqref{3.24}, \eqref{3.26},
\beq\label{11.5}
R_{0, A, B}(z)= M \,  R_{0, \tilde{A}, \tilde{B}}(z)\, M^\dagger, \quad z \in \rho \left( H_{0,A,B} \right).
\ene
Hence, by \eqref{3.20}-\eqref{3.22}and  \eqref{11.4} the integral kernel of  $R_{0, A, B}(z)$ is given by,
\beq \label{11.6}
R_{0, A, B}(z)(x,y)=\begin{cases}\varphi_{0,A, B}(x,k)\, e^{ik y}\, \left[J_{0, A,B}(k)\right]^{-1}, \quad x \leq y, \\
 e^{ik x}\,   \varphi_{0,A, B}(y,k) \,  \left[J_{0, A, B}(k)\right]^{-1}, \quad x \geq y. 
\end{cases}
\ene
The  estimate below follows from \eqref{3.29}, \eqref{3.32}, \eqref{11.4} and \eqref{11.5}
\begin{eqnarray}\label{11.6b}
\left|  R_{0, A, B}(z)(x,y) \right| \leq C  D(k) \, e^{-\textrm{Im} k |x-y|}, D(k):= \max \left[ \frac{1}{|k|},
\frac{1}{\left|\cos \theta_1+ik\sin\theta_1\right|},\dots,
\frac{1}{\left|\cos\theta_{n_{\mathrm M}}+ik\sin\theta_{n_{\mathrm M}}\right|}\right], k=\sqrt{z},  \nonumber \\ \label{11.7}\\
\nonumber
\left| \frac{\partial}{\partial x} R_{0, A, B}(z)(x,y) \right| \leq C  F(k)\, e^{-\textrm{Im} k |x-y|}, F(k):= \max \left[ 
\frac{1+|k|}{\left|\cos \theta_1+ik\sin\theta_1\right|},\dots,
\frac{1+|k|}{\left|\cos\theta_{n_{\mathrm M}}+ik\sin\theta_{n_{\mathrm M}}\right|}\right], k=\sqrt{z}.
\end{eqnarray}

It follows from \eqref{11.7} that  for $ z \in \mathbf C^\pm$, the resolvent $R_{0,A,B}(z)$ extends from $L^1 \cap L^2$ into a bounded operator in $\mathcal B\left( L^1, \mathbf{W}_{1,\infty} \right)$.
Furthermore, as
\beq\label{11.8}
 -\frac{d^2}{d \,x^2}\,R_{0,A,B}(z)= H_{0,A,B} \,R_{0,A,B}(z)=   I + z R_{0,A,B}(z),
 \ene
we have that $ R_{0,A,B}(z) \in \mathcal B\left( L^1, {\mathbf W}_{2,\infty}  \right)$, with norm uniformly bounded  for $z$ in any compact set of $\mathbf{C}^\pm$. Furthermore, by  \eqref{11.7} we also have that,  $ R_{0,A,B}(z) \in \mathcal B\left( L^1, L^1  \right)$, with norm uniformly bounded for $z$ in any compact set of $\mathbf{C}^\pm$. Then, for any $\psi \in L^1$,

$$
\frac{d}{d z}\,R_{0,A,B}(z)\, \psi= \left(R_{0,A,B}(z)\right)^2 \,\psi \in {\mathbf W}_{2,\infty}, \qquad \forall z \in \mathbf{C}^\pm,
$$
where the derivative with respect to $z$ exists in the strong topology of $W_{2,\infty}$.  Then, by Theorem 3.12 in page 152 of \cite{ka} $R_{0,A,B}(z)$ is analytic as an operator valued function from $\mathbf C^\pm$ into $\mathcal B\left( L^1, {\mathbf W}_{2,\infty} \right)$.

 In the following theorem we give a limiting absorption principle for $H_{0,A,B}$
\begin{theorem}\label{theo11.1}
For every $ \lambda >0$ the following limits,
\beq\label{11.9}
R_{0,A,B}(\lambda \pm i 0):= \lim_{\varepsilon \downarrow 0}\, R_{0,A,B}(\lambda\pm i \varepsilon),
\ene
 exist in the strong  operator topology in $ \mathcal B\left( L^1,  W_{2,\infty, } \right)$. Furthermore, the functions,
\beq \label{11.10}
R^{\pm}_{0,A,B}(\lambda):= \begin{cases} R_{0,A,B}(\lambda), \qquad \hbox{\rm if  } \lambda \in \mathbf C^\pm, \\
R_{0,A,B}(\lambda \pm i 0), \qquad \hbox{\rm  if }\lambda \in (0, \infty),\end{cases}
\ene
defined  for $\lambda \in \mathbf C^{\pm} \cup (0,\infty)$,  with values in  $ \mathcal B\left( L^1,  W_{2,\infty} \right)$ are analytic for $  \lambda \in \mathbf C^\pm$ and continuous in the strong operator topology for $\lambda \in  (0, \infty)$. 
\end{theorem}
                                                               
\noindent {\it Proof:} the theorem follows from \eqref{11.6}-\eqref{11.8}.

\bull

Note that it follows from \eqref{11.6} that the integral kernel of  $R_{0, A, B}(k^2 \pm i0)$ is given by,
\beq \label{11.11}
R_{0, A, B}(k^2\pm i0)(x,y)=\begin{cases} M \,\varphi_{0,\tilde{A}, \tilde{B}}(x,k)\, \left[J_{0, \tilde{A},\tilde{B}}( \pm k)\right]^{-1}\, e^{  \pm ik y} \,
M^\dagger, \quad x \leq y, \\
M \, e^{ \pm i  k x}\,  \left[J_{0, \tilde{A}, \tilde{B}}( \pm k)\right]^{-1}  \, \varphi_{0,\tilde{A}, \tilde{B}}(y,k) \, M^\dagger\, , \quad x \geq y. 
\end{cases}
\ene

\begin{remark}\label{rem11.2}{ \rm Equations  \eqref{11.7}, \eqref{11.8},  \eqref{11.11}  imply that,  for $ k^2 \in (0, \infty)$},
\beq\label{11.12}
\left\| R^\pm_{0,A,B}(k^2)\right\|_{\mathcal B \left(  L^1, L^{\infty}\right)} \,\leq \frac{C}{|k|},
\left\| R^\pm_{0,A,B}(\lambda)\right\|_{\mathcal B \left(  L^1, W_{1,\infty}\right)} \leq C\, \frac{1+|k|}{|k|}, 
\left\| R^\pm_{0,A,B}(\lambda)\right\|_{\mathcal B \left(  L^1, W_{2,\infty}\right)} \leq C  \, \frac{1+|k|^2}{|k|}. 
\end{equation}
\end{remark}

\bull

\subsection{The case of integrable potential}

Let,
\beq\label{11.13}
V= \hat{U} |V|,
\ene
be the polar decomposition of $V$ \cite{ka}, with $\hat{U}$ partially isometric, and $|V|$ the absolute value of $V$.  We have that, $\hat{U} =\hat{U}^*, \hat{U} |V|= |V| \hat{U}$.  Denote,
\beq\label{11.14}
V_1:= \sqrt{|V|}, \qquad V_2:= \hat{U} \sqrt{|V|}.
\ene

Since, $ V_j \in L^2, j=1,2$  and $ \mathbf{H}_1 \subset L^\infty$ we have that,  $V_j \in \mathcal B \left( \mathbf{H}_1, L^2   \right)$ and also by duality,    $V_j \in \mathcal B \left( L^2, \mathbf{H}_{-1} \right)$.  Then, the quadratic form \eqref{10.11} can be written as,
\beq \label{11.15}
h_{A,B}\left(\varphi,\psi\right):= h_{0,A,B}\left( \varphi,\psi\right)+ \left(V_1\,\varphi, V_2\psi\right), \quad D(h_{A,B})= \mathbf{H}_{1,A,B}.
\ene
 
 Let us denote by $R_{A,B}(z):= \left( H_{A,B} -z \right)^{-1}$ the resolvent of $H_{A,B}$ for $z \in \rho\left(H_{A,B}\right).$ By \eqref{11.15} with $\varphi= R_{0,A,B}(z) f, \psi= R_{A,B}(\overline{z})g$ we obtain that,
 
 \beq\label{11.16}
  R_{A,B}(z)- R_{0,A,B}(z)= - R_{A,B}(z)\, V_2\, V_1 R_{0,A,B}(z)= - R_{0,A,B}(z)\, V_2\, V_1 R_{A,B}(z), \quad z \in \rho(H_{0,A,B})\cap \rho(H_{A,b}).
 \ene
 
 As $D(H_{0,AB}) \subset \mathbf{H}_1, D(H_{A,B} \subset \mathbf{H}_1$, it follows that, $R_{0,A,B}(z) \in \mathcal B \left( L^2, \mathbf{H}_1 \right)$,  $ R_{A,B}(z) \in \mathcal B \left( L^2, \mathbf{H}_1 \right)$ and by   duality  $ R_{0,A,B}(z) \in \mathcal B \left( \mathbf{H}_{-1}, L^2 \right)$,  $R_{A,B}(z) \in \mathcal B \left(\mathbf{H}_{-1}, L^2\right)$.  
 
By \eqref{11.7}  $V_1 R^{\pm}_{0,A,B}(z) V_2$ is a Hilbert-Schmidt operator in $L^2$ and it Hilbert-Schmidt norm  satisfies,
\beq \label{11.17}
\left\| V_1 R^{\pm}_{0,A,B}(z)) V_2 \right\|_{\textrm{HS}}\, \leq  C\, D(k),\, D(k):= \max \left[ \frac{1}{|k|},
\frac{1}{\left|\cos \theta_1+ik\sin\theta_1\right|},\dots,
\frac{1}{\left|\cos\theta_{n_{\mathrm M}}+ik\sin\theta_{n_{\mathrm M}}\right|}\right],\, k= \sqrt{z}.
\ene
Hence, by \eqref{11.16}

 \beq\label{11.18}
 \left(I- V_1\,R_{A,B}(z)\, V_2\right)\, \left( I+ V_1 R_{0,A,B}(z)\, V_2 \right) \, \varphi= \varphi, \qquad \forall \varphi \in L^2.
 \ene
In Lemma \ref{lemm11.3} below we prove that $\left(I+ V_1\,R_{0,A,B}(z)\, V_2\right)$ is invertible in $L^2$. Then, applying  \eqref{11.18} to\linebreak $\varphi= \left(I+ V_1 R_{0,A,B}(z) V_2 \right)^{-1} \psi$ we obtain,
 
 \beq \label{11.19} 
  \left(I- V_1\,R_{A,B}(z)\, V_2\right) \, \psi=\left(I+ V_1 R_{0,A,B}(z)\right)^{-1} \psi, \qquad \forall \psi \in L^2.
  \ene
  
It follows that, $ V_1\,R_{A,B}(z)\, V_2 \in \mathcal B\left(  L^2   \right)$ and
\beq \label{11.20}
I- V_1\,R_{A,B}(z)\, V_2 =\left(I+ V_1 R_{0,A,B}(z) V_2\right)^{-1}.
\ene

Iterating \eqref{11.16} and using \eqref{11.19} we prove that,
\beq\label{11.21} 
R_{A,B}(z)= R_{0,A,B}(z)- R_{0,A,B}(z)\, V_2  \,\left(I+ V_1 R_{0,A,B}(z)\, V_2\right)^{-1} \, V_1 R_{0,A,B}(z),
\quad z \in \rho(H_{0,A,B})\cap \rho(H_{A,b}).
\ene
 Denote,
 \beq \label{11.22}
 Q_{\pm}(\lambda)=  V_1\, R^\pm_{0,A,B}(\lambda) \, V_2, \qquad \lambda \in \mathbf C^\pm \cup (0, \infty).
\ene
\begin{lemma}\label{lemm11.3}
Suppose that $V$ satisfies \eqref{1.6}, \eqref{1.7}. Then, for every $ \lambda \in \mathbf C^\pm \cup (0,\infty)$ the operator $I + Q_{\pm}(\lambda)$ is invertible in $L^2$.
\end{lemma}
\noindent {\it Proof:} since $Q_{\pm}(\lambda)$ is compact we only have to prove that $-1$ is not an eigenvalue. On the contrary, suppose that there is a $\varphi \in L^2$ such that,

$$
\varphi= - Q_{\pm}(\lambda)\, \varphi.
$$
 Denote,
 $$
 \psi:=  R^\pm_{0,A,B}(\lambda)\, V_2 \,\varphi.
 $$ 
 Then, $\psi \in L^\infty$, it satisfies the boundary condition \eqref{1.2} and
 $$
 H_{A,B}\, \psi = \lambda\, \psi. 
 $$
  Moreover, $ \psi$ can be written as in \eqref{3.7} but with $\eta =0$ because it is bounded. Then, if $ \lambda \in {\mathbf C}^\pm,  \psi \in L^2$ and $ \lambda$ would be a complex eigenvalue of $H_{A,B}$, what is impossible because $H_{A,B}$ is selfadjoint. If $ \lambda \in (0,\infty)$, it follows from \eqref{11.11} that,
    
   $$
  \psi = e^{\pm i k x}\, \xi + o(1), \,\, x \rightarrow \infty, \,\, \,\textrm{for some}\, \xi \in \mathbf {C}^n, k=\sqrt{\lambda}. 
  $$
But then,
$$
\psi- f(\pm k, x)\, \xi \rightarrow 0,\,\, x \rightarrow \infty.
$$
However, as $ \psi - f(\pm k, x) \, \xi$ is a solution to \eqref{3.1}, it is a linear combination of $f(k,x)$ and $g(k,x)$, but as $f(k, x)$ behaves as $ e^{ i kx}$  and $g(k,x)$ as $e^{-ik x},   x \rightarrow \infty$ (see \eqref{3.2}, \eqref{3.5}),  necessarily, $ \psi - f(\pm k, x)\, \xi=0$. Hence,
$$
\psi = f (\pm k,x) \xi,
$$
and as $ \psi$ satisfies the boundary condition \eqref{1.2}, it follows that,
$$
0= \left(- B^\dagger\, f(\pm k,0) + A^\dagger\, f'(\pm k, 0)\right)\, \xi = -J^\dagger(\mp k)\, \xi.
$$
This implies that $ \xi=0$ because $J^\dagger(k)$ is invertible for $  k \in \mathbb R\setminus 0$ \cite{19}. It follows that $ \psi =0$.

\bull

We give below a limiting absorption principle for $H_{A,B}$.
\begin{theorem}\label{theo11.4}
Suppose that $V$ satisfies \eqref{1.6}, \eqref{1.7}. Then, for every $ \lambda >0$ the following limits,
\beq\label{11.23}
R_{A,B}(\lambda \pm i 0):= \lim_{\varepsilon \downarrow 0}\, R_{A,B}(\lambda\pm i \varepsilon),
\ene
 exist in the strong  operator topology in $ \mathcal B\left( L^1,  W_{2,\infty, } \right)$. Furthermore, the functions,
\beq \label{11.24}
R^{\pm}_{A,B}(\lambda):= \begin{cases} R_{A,B}(\lambda), \qquad \hbox{\rm if  } \lambda \in \mathbf C^\pm, \\
R_{A,B}(\lambda \pm i 0), \qquad \hbox{\rm  if }\lambda \in (0, \infty),\end{cases}
\ene
defined  for $\lambda \in \mathbf C^{\pm} \cup (0,\infty)$,  with values in  $ \mathcal B\left( L^1,  W_{2,\infty,} \right)$ are analytic for $  \lambda \in \mathbf C^\pm$ and  continuous in the strong operator topology for $\lambda \in  (0, \infty)$. 
\end{theorem}
                                                               
\noindent {\it Proof:} the theorem follows from Theorem \ref{theo11.1}, Lemma \ref{lemm11.3} and \eqref{11.21}.   
\bull

Note that,
\beq\label{11.25} 
R^\pm_{A,B}(\lambda)= R^\pm_{0,A,B}(\lambda)- R^\pm_{0,A,B}(\lambda)\, V_2  (I+ Q_{\pm}(\lambda))^{-1} \, V_1 R^\pm_{0,A,B}(\lambda), \qquad \lambda \in \mathbf C^\pm \cup (0,\infty).
\ene

Moreover, by  \eqref{11.12},  \eqref{11.17} and  \eqref{11.25}, for every $ \delta >0$ there is a constant $C_\delta$
such that  for $ k^2 \in ( \delta^2, \infty)$,
\beq\label{11.26}
\left\| R^\pm_{A,B}(k^2)\right\|_{\mathcal B \left(  L^1, L^{\infty}\right)} \,\leq \frac{C_\delta}{|k|},
\left\| R^\pm_{0,A,B}(k^2)\right\|_{\mathcal B \left(  L^1, W_{1,\infty}\right)} \leq C_\delta\, \, 
\left\| R^\pm_{0,A,B}(k^2)\right\|_{\mathcal B \left(  L^1, W_{2,\infty}\right)} \leq C_\delta  \, |k|. 
\end{equation}

\section{The generalized Fourier maps  } \sss
\subsection{The case of zero potential}
 It is proven in  the comments at the beginning of Section VIII of \cite{21} that $H_{0,A, B}$ has no positive or zero eigenvalues. For $ \lambda \in \mathbb R$ we denote by $E_{0,A,B}(\lambda )$ the spectral family of $H_{0,A,B}$ and for  any Borel set $O$ we designate by $E_{0,A, B} (O)$ the  spectral projector of $H_{0,A, B}$ for $O$. 
 
We first consider the case of the representation where the boundary conditions \eqref{1.2} are given by the real diagonal matrices $\tilde{A}, \tilde{B}$ \eqref{3.25}.  By Stone's formula, Theorem \ref{theo11.1} and Remark\ref{rem11.2},    for all $ 0 < a < b$,
\begin{eqnarray}\label{12.1}
\left( E_{0,\tilde{A}, \tilde{B}}((a,b))\, \varphi, \psi \right)&=& \lim_{\varepsilon \downarrow 0}\, \frac{1}{2\pi i}\, \int_a^b\, d \lambda \left( \left[ R_{0,\tilde{A}, \tilde{B}}(\lambda +i \varepsilon)- R_{0,\tilde{A}, \tilde{B}}(\lambda -i \varepsilon)\right]\varphi, \psi   \right)\\ \nonumber
&=&  \frac{1}{2\pi i}\, \int_a^b\,\, d \lambda \left( \left[R_{0,\tilde{A}, \tilde{B}}(\lambda +i 0)- R_{0,\tilde{A}, \tilde{B}}(\lambda -i 0)\right]\, \varphi , \psi \right) \qquad  \forall  \varphi, \psi \in L^2 \cap L^1.
\end{eqnarray}

Hence, the spectrum of $H_{0,\tilde{A}, \tilde{B}}$ in $(0, \infty)$ is absolutely continuous and
\beq \label{12.2}
\frac{d}{d \lambda}\,\left( E_{0,\tilde{A}, \tilde{B}}(\lambda)\, \varphi, \psi \right)=\frac{1}{2 \pi i}\,\left( \left[R_{0,\tilde{A}, \tilde{B}}(\lambda +i 0)- R_{0,\tilde{A}, \tilde{B}}(\lambda -i 0)\right]\, \varphi , \psi \right), \, \lambda >0, \qquad \forall  \varphi, \psi \in L^2 \cap L^1.
\ene

Let us denote by $\psi_{0,\tilde{A}, \tilde{B}}(k,x)$ the physical solution \eqref{3.13} with $ A=\tilde{A}, B=\tilde{B}$ and $V=0$.
 \beq \label{12.2b}
\psi_{0,\tilde{A},\tilde{B}}(k,x):= - i k \, \varphi_{0, \tilde{A}, \tilde{B}}(k,x)\, J^{-1}_{0,\tilde{A},\tilde{ B}}(k).
\ene

Let  $  E_{0,\tilde{A}, \tilde{B}}(O,x,y)$ be the integral kernel of   
$ E_{0,\tilde{A}, \tilde{B}}(O)$. Then, by \eqref{11.11}  and \eqref{3.12} with  $ A=\tilde{A}, B=\tilde{B}$, $M=I$ and \eqref{12.1},

\beq \label{12.4}
E_{0,\tilde{A}, \tilde{B}}((k_0^2, k_1^2)(x,y)=\, \frac{2}{\pi}\, \int_{k_0}^{k_1}\, \psi_{0,\tilde{A}, \tilde{B}}(k,x)\,  \psi^\dagger_{0,\tilde{A}, \tilde{B}}(k,y)\,\, dk,
\ene
where we also used that $J^\dagger_{0,\tilde{A}, \tilde{B}}(k)=J_{0,\tilde{A}, \tilde{B}}(-k)$, see \eqref{3.29}.

Let us define,
\beq\label{12.5}
\psi_{0, \tilde{A}, \tilde{B}}^{+}(k,x):= \psi_{0, \tilde{A}, \tilde{B}}(-k,x)  , \qquad \psi_{0, \tilde{A}, \tilde{B}}^{-}(k,x):= \psi_{0, \tilde{A}, \tilde{B}}(k,x).
\ene
Note that,
 $$
 \psi_{0, \tilde{A}, \tilde{B}}^{+}(k,x)= \overline{ \psi_{0, \tilde{A}, \tilde{B}}(k,x)}.
 $$ 
By  \eqref{3.29}, \eqref{3.31}, \eqref{3.32}, \eqref{12.2b} and \eqref{12.5},
\beq \label{12.5b}
\psi_{0, \tilde{A}, \tilde{B}}^{\pm }= \mathrm{diag}\{ \psi_{0, \tilde{A}, \tilde{B},1}^{\pm }, \psi_{0, \tilde{A}, \tilde{B},2}^{\pm }, \cdots ,\psi_{0, \tilde{A}, \tilde{B},n}^{\pm } \},
\ene

\beq\label{12.5c}
\psi_{0, \tilde{A}, \tilde{B},j}^{\pm }(k,x)= \begin{cases} \displaystyle \pm i\, \sin( kx),  \,\textrm{if}\, \theta_j=\pi, \\
\displaystyle \cos kx, \, \textrm{if}\, \theta_j= \pi/2, \\
\displaystyle   i \left[ \pm \cos\theta_j \, \sin kx \mp k \sin\theta_j\, \cos kx  \right] \frac{1}{ \cos\theta_j \mp i k \sin\theta_j} \,\theta_j \neq \pi, \pi/2.
\end{cases}
\ene

We denote by $F_{0,\tilde{A},\tilde{B}}^\pm$ the generalized Fourier maps,

\beq\label{12.6}
\left(F_{0,\tilde{A},\tilde{B}}^\pm \psi \right)(k):= \sqrt{\frac{2}{\pi}}\, \int_0^\infty\, \left(\psi^\pm_{0,\tilde{A}, \tilde{B}}(k,x)\right)^\dagger\, \psi(x)\, dx, \qquad \psi \in L^2\cap L^1.
\ene
Then, by \eqref{12.4}, denoting by $\left<  \cdot, \cdot \right>$ the scalar product in $\mathbf C^n$,
 \beq \label{12.7}
 \left( E_{0,\tilde{A}, \tilde{B}}((k_0^2, k_1^2))\psi_1, \psi_2\right)= \int_{k_0}^{k_1}\,   \left< \left(F_{0,\tilde{A},\tilde{B}}^\pm \psi_1 \right)(k)\,,  \left(F_{0,\tilde{A},\tilde{B}}^\pm \psi_2 \right)(k)\right>\,dk, \qquad \forall \psi_1, \psi_2 \in L^2 \cap L^1,
\ene

where we used that $\psi^\pm_{0,\tilde{A}, \tilde{B}}(k,x)$ are diagonal matrices and that $\varphi_{0,\tilde{A}, \tilde{B}}(x,k)$ is real valued for $k$ real. Taking the limit $ k_0 \rightarrow 0, k_1 \rightarrow \infty$  with $\psi_1=\psi_2$, we obtain that,

\beq \label{12.8}
\left\| E_{0,\tilde{A}, \tilde{B}}((0, \infty))  \psi \right\|= \left\| F_{0,\tilde{A},\tilde{B}}^\pm \psi \right\|_{L^2}, \qquad \forall \psi \in  L^2 \cap L^1.
\ene
Hence, the $ F_{0,\tilde{A},\tilde{B}}^\pm$ extend to bounded operators on $L^2$.

In the next theorem we give the properties of the generalized Fourier maps,   $F_{0,\tilde{A},\tilde{B}}^\pm $.

\begin{theorem} \label{theo12.1}
The Hamiltonian $H_{0, \tilde{A}, \tilde{B}}$ has no positive or zero eigenvalues. Its negative spectrum consists of a finite number of  eigenvalues of multiplicity smaller or equal to $n$. The non-negative spectrum is $[0, \infty)$ and it is absolutely continuos. The generalized Fourier maps  $F_{0,\tilde{A},\tilde{B}}^\pm $ are partially isometric  with initial subspace $ \mathcal{ H}_{ \textrm ac}(H_{0, \tilde{A}, \tilde{B}})$ and final  subspace  $L^2$. Moreover, the adjoint operators are given by,
\beq \label{12.9}
\left( \left(F_{0,\tilde{A},\tilde{B}}^\pm \right)^\dagger \psi \right)(x)= \, \sqrt{\frac{2}{\pi}}\, \int_0^\infty\, \left(\psi^\pm_{0,\tilde{A}, \tilde{B}}(k,x)\right)\, \psi(k)\, dk, \qquad \psi \in L^2 \cap L^1.
\ene
Furthermore,
\beq \label{12.10}
 F_{0,\tilde{A},\tilde{B}}^\pm\, H_{0,\tilde{A}, \tilde{B}}\,\left(F_{0,\tilde{A},\tilde{B}}^\pm \right)^\dagger= k^2.
 \ene 
\end{theorem}
 \noindent {\it Proof:}  Recall that it is proven in  the comments at the beginning of Section VIII of \cite{21} that $H_{0,\tilde{A}, \tilde{B}}$ has no positive or zero eigenvalues.  Furthermore in  Theorems 8.1 and 8.6 of \cite{21} ) it is established that it  has a finite number of negative eigenvalues of  multiplicity less or equal  to $n$, that coincide with the points $-k^2$ such that $J_{0, \tilde{A}, \tilde{B}}(i k)$ is not invertible for $ k \in \mathbf{R}^+$. Let us  denote by $\sigma_{0,p} $ the set of negative eigenvalues of $H_{0,\tilde{A}, \tilde{B}}$. Hence, by \eqref{11.4}, ]\eqref{11.7} the resolvent $R_{0, \tilde{A},\tilde{B}}(z)$ is a bounded operator in $L^2$ for $ z \in (-\infty, 0) \setminus \sigma_{0,p}$, and in consequence the only spectrum in $(-\infty, 0)$ is $ \sigma_{0,p}$. By \eqref{12.2} $ H_{0, \tilde{A},\tilde{B}}$ has no singular continuous spectrum.  It follows from \eqref{12.8} that $F_{0,\tilde{A},\tilde{B}}^\pm$ are partially isometric with initial subspace  $E_{0,\tilde{A}, \tilde{B}}((0, \infty))L^2$   into $L^2$. Equation \eqref{12.9} is immediate from the definition of  $F_{0,\tilde{A},\tilde{B}}^\pm$ in \eqref{12.6}, and \eqref{12.10} follows from the fact that   $ \psi^\pm_{0,\tilde{A}, \tilde{B}}(k,x)$ are solution to the Schr\"odinger equation \eqref{3.1} But, then by \eqref{12.10} $[0,\infty)$ actually belongs to the spectrum of $H_{0,\tilde{A},\tilde{B}}$ and  $\mathcal{H}_{\textrm ac}(H_{0,\tilde{A}, \tilde{B}})= E_{0,\tilde{A}, \tilde{B}}((0, \infty))L^2$.
 
 It remains to prove that the   $F_{0,\tilde{A},\tilde{B}}^\pm$ are onto $L^2$. Suppose, on the contrary, that there is $ \phi_\pm \in L^2$ such that,
 
 \beq\label{12.11}
 \left( \phi_\pm, F_{0,\tilde{A},\tilde{B}}^\pm\, E_{0,\tilde{A}, \tilde{B}}(k_0^2,k_1^2)\, \psi\right)=0, \qquad \forall \, \psi \in L^2\cap L^1, \forall \, 0 < k_0 , k_1. 
 \ene
Since by \eqref{12.7},
\beq\label{12.12}
\chi_{(k_0, k_1)}(k)  F_{0,\tilde{A},\tilde{B}}^\pm  =  F_{0,\tilde{A},\tilde{B}}^\pm \, E_{0,\tilde{A}, \tilde{B}}(k_0^2, k_1^2),
\ene
where $\chi_{(k_0, k_1)}$ is the characteristic function of $(k_0, k_1)$, it follows from \eqref{12.11} that,

\beq \label{12.12b}
  \left(F_{0,\tilde{A},\tilde{B}}^\pm\,\right)^\dagger\,  \chi_{(k_0, k_1)} \phi_\pm =\sqrt{\frac{2}{\pi}} \,\int_{k_0}^{k_1} \, \psi^\pm_{0,\tilde{A}, \tilde{B}}(k,x)\, \phi_\pm(k)\, dk=0, \qquad \forall \,0 <k_0 < k_1.
\ene
 Since $ \varphi_{0,\tilde{A}, \tilde{B}}(k,0)= \tilde{A},  \varphi'_{0,\tilde{A}, \tilde{B}}(k,0)= \tilde{B}$ it follows from 
  \eqref{12.2b}, \eqref{12.5} and \eqref{12.12b} that,
  
  $$
  \int_{k_0}^{k_1} \, k\,  (\tilde{B} + i\tilde{A}) \,\, \left(J_{0,\tilde{A}, \tilde{B}}(\mp k)\right)^{-1} \phi_\pm (k)\, dk=0, \qquad \forall \, 0 <k_0 < k_1.
$$
 Finally since $(\tilde{B}+i \tilde{A})\, \left(J_{0,\tilde{A}, \tilde{B}}(\mp k)\right)^{-1}$ is invertible for $ k \neq 0$ (see \eqref{3.25}, \eqref{3.29}) we obtain that $ \phi_\pm =0$.

 \bull
 
 Let us now consider the case of $A,B$. We denote by $ \psi_{0,A,B}(k,x)$    the physical solution \eqref{3.13} in the case $V=0,$
 
  \beq \label{12.13}
\psi_{0,A, B}(k,x):= - i k \, \varphi_{0,A, B}(k,x)\, J^{-1}_{0,A, B}(k).
\ene

We define,
\beq \label{12.14} 
\psi_{0, A, B}^{\pm}(k,x):= \psi_{0,A,B}(\mp k,x).
\ene
By  \eqref{3.21} and \eqref{3.22},
\beq\label{12.15}
\psi_{0,A, B}^\pm(k,x):=  M\, \psi_{0,\tilde{A}, \tilde{B}}^\pm(k,x)\, M^\dagger.
\ene  
We define the generalized Fourier maps for $A_{0,A,B}$ as,
\beq \label{12.16}
\left(F_{0,A, B}^\pm \psi \right)(k):= \sqrt{\frac{2}{\pi}}\, \int_0^\infty\, \left(\psi^\pm_{0,A, B}(k,x)\right)^\dagger\, \psi(x)\, dx, \qquad \psi \in L^2\cap L^1.
\ene
It follows from \eqref{12.15} that,
\beq\label{12.17}
F_{0,A, B}^\pm = M\, F_{0,\tilde{A}, \tilde{B}}^\pm  \, M^\dagger.
\ene

Hence, by \eqref{3.24}, \eqref{3.26}, \eqref{12.7} and \eqref{12.17},
 \beq \label{12.18}
 \left( E_{0,A,B}((k_0^2, k_1^2))\psi_1, \psi_2\right)= \int_{k_0}^{k_1}\,   \left<\left(F_{0,A, B}^\pm \psi_1 \right)(k),\,  \left(F_{0,A,B}^\pm \psi_2 \right)(k)\right>\,dk, \qquad \forall \psi_1, \psi_2 \in L^2 \cap L^1.
\ene

\begin{theorem} \label{theo12.2}
The Hamiltonian $H_{0, A, B}$ has no positive  eigenvalues. Its negative spectrum consists of a finite number of  eigenvalues of multiplicity smaller or equal to $n$. The non negative spectrum is $[0, \infty)$ and it is absolutely continuos. The generalized Fourier maps  $F_{0,A,B}^\pm $ partially isometric   with initial subspace $ \mathcal{ H}_{ \textrm ac}(H_{0, A, B})$ and final  subspace  $L^2$. Moreover, the adjoint operators are given by,

\beq \label{12.19}
\left( \left(F_{0,A,B}^\pm \right)^\dagger \psi \right)(x)= \, \sqrt{\frac{2}{\pi}}\, \int_0^\infty\, \left(\psi^\pm_{0,A, B}(k,x)\right)\, \psi(k)\, dk, \qquad \psi \in L^2 \cap L^1.
\ene
Furthermore,
\beq \label{12.20}
 F_{0,A,B}^\pm\, H_{0,A, B}\,\left(F_{0,A,B}^\pm \right)^\dagger= k^2.
 \ene 
\end{theorem}

\noindent{\it Proof:} The theorem follows from \eqref{3.24}, \eqref{3.26}, Theorem \ref{theo12.1} and \eqref{12.17}.

\bull
\subsection{The case of integrable potential.}
 It is proven in  the comments at the beginning of Section VIII of \cite{21} that $H_{A, B}$ has no positive  eigenvalues.
We denote,
\beq \label{12.21}
L^\pm_{A,B} (\lambda) := I- V\, R^\pm_{A,B} (\lambda), \qquad \lambda \in \mathbf C^\pm \cup (0,\infty).
\ene
Then, by  Theorem \ref{theo11.1} and \eqref{11.16},
\beq \label{12.22}
R^\pm_{A,B}(\lambda)= R^\pm_{0,A,B}(\lambda)\, L^\pm_{A,B}(\lambda), \qquad \lambda \in \mathbf C^\pm \cup (0,\infty).
\ene
Since,
$$
R_{A,B}(z)- R_{A,B}(\overline{z})= (z- \overline{z})\, R_{A,B}(z)\, R_{A,B}(\overline{z}), \qquad z \in \mathbf C^\pm,
$$
Then,
\beq \label{12.23}
\frac{1}{2\pi i}\left[ R_{A,B}(z)-R_{A,B}(\overline{z}) \right]= \left(L^+_{A,B}\right)^\star (z)\, \frac{1}{2\pi i}\left[ R_{0,A,B}(z)-R_{0,A,B}(\overline{z}) \right] \, L^+_{A,B}(z), \qquad z \in \mathbf C^+,
\ene
and  also,
$$
R_{A,B}(z)- R_{A,B}(\overline{z})= (z- \overline{z})\, R_{A,B}(\overline{z})\, R_{A,B}(z), \qquad z \in \mathbf C^\pm,
$$
we have that,
\beq \label{12.24}
\frac{1}{2\pi i}\left[ R_{A,B}(z)-R_{A,B}(\overline{z}) \right]= \left(L^-_{A,B}\right)^\star (\overline{z})\, \frac{1}{2\pi i}\left[ R_{0,A,B}(z)-R_{0,A,B}(\overline{z}) \right] \, L^-_{A,B}(\overline{z}), \qquad z \in \mathbf C^+.
\ene

We define the generalized Fourier maps for $H_{A,B}$ as follows,

\beq \label{12.25}
\left(F_{A, B}^\pm \psi \right)(k):= \sqrt{\frac{2}{\pi}}\, \int_0^\infty\, \left(\psi^\pm_{0,A, B}(k,x)\right)^\dagger\, L^\pm(k^2)\psi(x)\, dx, \qquad \psi \in L^2\cap L^1.
\ene

For  $ \lambda \in \mathbb R$ we denote by   $E_{A,B}(\lambda)$ the spectral family of $H_{A,B}$ and for any Borel set $O$ let us designate by $E_{A,B}(O)$ the spectral projector of $H_{A,B}$ for $O$. Then, by Theorems \ref{theo11.4}, \ref{theo12.1}, \eqref{12.18}, \eqref{12.23}, \eqref{12.24} and Stone's formula we have that,

\beq \label{12.26}
 \left( E_{A,B}((k_0^2, k_1^2))\psi_1, \psi_2\right)= \int_{k_0}^{k_1}\,   \left<\left(F_{A, B}^\pm\,  \psi_1 \right)(k),\,  \left(F_{A,B}^\pm \psi_2 \right)(k)\right>\,dk, \qquad \forall \psi_1, \psi_2 \in L^2 \cap L^1, \forall \,0 < k_0 < k_1,
\ene
where we used that $H_{A,B}$ has no positive  eigenvalues.

Taking the limit $ k_0 \rightarrow 0, k_1 \rightarrow \infty$ in \eqref{12.26}  with $ \psi_1= \psi_2= \psi$ we get that,
\beq\label{12.27}
\left\|  F^\pm_{A,B} \psi \right\|_{L^2}= \left\| E_{A,B}([0, \infty)) \psi \right\|_{L^2}.
\ene 
Then, the $F^\pm_{A,B}$ extend to bounded operators
 in $L^2$.
\begin{theorem} \label{theo12.3} Suppose that $ V$ satisfies \eqref{1.6}, \eqref{1.7}. Then, $H_{A,B}$ has no  positive eigenvalues, and the negative spectrum of $H_{A,B}$ consists of isolated eigenvalues of  multiplicity  smaller or equal than $n$, that can accumulate only at zero. Furthermore, $H_{A,B}$ has no singular continuous spectrum and its absolutely continuous spectrum is given by $[0, \infty)$. 
The generalized Fourier maps  $F_{A,B}^\pm $ are partially isometric   with initial subspace 
$ \mathcal H_{\textrm {ac}}(H_{A,B})$ and final subspace  $L^2$. Moreover, the adjoint operators are given by,
\beq \label{12.28}
\left( \left(F_{A,B}^\pm \right)^\dagger \psi \right)(x)= \, \sqrt{\frac{2}{\pi}}\, \int_0^\infty\, \left( L^\pm(k^2) \right)^\dagger \,    \psi^\pm_{0,A, B}(k,x)\, \, \psi(k)\, dk, \qquad \psi \in L^2 \cap L^1.
\ene
Furthermore,
\beq \label{12.29}
 F_{A,B}^\pm\, H_{A, B}\,\left(F_{A,B}^\pm \right)^\dagger= k^2.
 \ene 
\end{theorem}

\noindent{Proof:} Recall that it is proven in  the comments at the beginning of Section VIII of \cite{21} that $H_{A, B}$ has no positive  eigenvalues.
 In Lemma \ref{lemm15.1} in Section ~9 we prove that for all $z \in \rho\left( H_{0,A,B} \right) \cap  \rho\left( H_{A,B} \right)$ the difference of the resolvents of $H_{0,A,B}$ and $H_{A,B}$ is trace class. Then, the essential spectrum of $H_{A,B}$ is $[0, \infty)$ and the negative spectrum consists of isolated eigenvalues of finite  multiplicity  that can accumulate only at zero. It  follows from item (b) of Theorem 8.1 in \cite{21} that the multiplicities of the eigenvalues is smaller or equal to $n$. Note that in Section~ 8 of \cite{21} the condition that the potential has a finite first moment is only used in order that the number of eigenvalues is finite. By \eqref{12.26} $H_{A,B}$ has no singular continuous spectrum. The fact that the $F^\pm_{A,B}$ are partially isometric with initial subspace $\mathcal H_{\textrm ac}(H_{A,B})$ into $L^2$ and \eqref{12.29} follow from \eqref{12.26} and \eqref{12.27}.  Equation \eqref{12.28} follows from the definition of $F^\pm_{A,B}$ in \eqref{12.25}. Equation \eqref{12.29} is implied by \eqref{12.26}.
 
  It remains to prove that the $F^\pm_{A,B}$ are onto $L^2$.

By \eqref{11.16} and Theorems \ref{theo11.1}, and \ref{theo11.4},
\beq\label{12.30}
L^\pm(\lambda)\left( I+ V\,R^\pm_{0,A,B}(\lambda)\right)=I,  \qquad \forall \lambda \in \mathbf C^\pm \cup (0,\infty),
\ene
as bounded operators on $L^1$. Let us prove that $V \,R^\pm_{0,A,B}(\lambda)$ is compact in $L^1$. Let $\psi_n$ be a bounded sequence in $L^1$. Since $R^\pm_{0,A,B}(\lambda) \in \mathcal B \left( L^1, W_{2, \infty}\right)$ the sequence 
  $R^\pm_{0,A,B}(\lambda) \psi_n$ is uniformly equicontinuous in $[0, R]$ for any $R >0$ Then, by the Arzel\'a-Ascoli theorem  there is a subsequence $\psi_{n_l}$ of $\psi_n$ such that  $R^\pm_{0,A,B}(\lambda) \psi_{n_l}$ converges uniformly  in $C([0,R]$ and then,    $ \chi_{[0,R]}(x) V(x) \, R^\pm_{0,A,B}(\lambda) \psi_{n_l}$ converges in $L^1$. Hence the operator $\chi_{[0,R]}(x) \, V(x)\, R^\pm_{0,A,B}(\lambda)$ is compact, but since
  
  $$
  \left\| V(x)\,   R^\pm_{0,A,B}(\lambda)- \chi_{[0,R]}(x) \, V(x)\, R^\pm_{0,A,B}(\lambda)\right\|_{\mathcal B(L^1)} \leq 
  C \,\left\| \chi_{(R, \infty)}(x) V(x)\right\|_{L^1} \rightarrow 0, \,\textrm{as} \, R \rightarrow \infty,
  $$ 
 it follows that $  V(x)\,   R^\pm_{0,A,B}(\lambda) $ is compact. Suppose that for some $ \lambda \in \mathbf C^\pm \cup (0,\infty)$,  $\left( I+ V  \, R^\pm_{0,A,B}(\lambda)  \right)$  is not injective in $L^1$. Then, by duality, also  $\left( I +  \, R^\mp_{0,A,B}(\overline{\lambda}) \,V  \right)$ is not injective in $L^\infty$. In consequence, there is a $\psi \in L^\infty$ such that,
 
 \beq \label{12.31}
 \psi = -  R^\mp_{0,A,B}(\overline{\lambda}) \,V  \,  \psi.
 \ene
 But then, $\phi:= V_1 \psi \in L^2$, satisfies  (recall that $V_1\, V_2 = V_2\, V_1 = V$ ),
 $$
 \phi = - Q_{\mp}(\overline{\lambda})\, \phi.
 $$
 By Lemma \ref{lemm11.3}, $\phi=0$ and then, by \eqref{12.31} $\psi=0$. Hence,  $\left( I+ V  \, R^\pm_{0,A,B}(\lambda)  \right)$ have a bounded inverse in $L^1$ and by \eqref{12.30} the $L^\pm(\lambda)$ are bounded invertible in $L^1$ and
 \beq \label{12.32}
 \left(L^\pm(\lambda)\right)^{-1}= \left( I+ V  \, R^\pm_{0,A,B}(\lambda)  \right)^{-1}, \qquad \forall \lambda \in \mathbf C^\pm \cup (0,\infty).
 \ene
It follows that $L^\pm(\lambda)$ are injective and onto $L^2$.

Suppose that $\phi$ is orthogonal to the range of $F^\pm_{A,B}$.  Let $\{\psi_n\}_{n=1}^\infty$ be a set of functions in $L^2\cap L^1$ that is countable dense in $L^1$. Then for all $ 0 < k_0 < k_1$ $$
\left( \phi, F^\pm_{A,B} E_{A,B}(k_0^2, k_1^2) \psi_n\right) = \sqrt{\frac{2}{\pi}}\, \int_{k_0}^{k_1} \, dk \, \left( \, \psi^\pm_{0, A,B}(k, \cdot)\, \phi(k), \, L^\pm(k^2)\,   \psi_n \right)=0.    
$$  
It follows that there is a set of measure zero, $ O$, independent of $n$  such that,
$$
 \left(\psi^\pm_{0, A,B}(k, \cdot)\, \phi(k), L^\pm(k^2)\, \psi_n \right)=0, \forall k \in (0,\infty)\setminus O.
$$
But since $L^\pm(k^2)$ is onto, also $\left\{L^\pm(k^2)\, \psi_n\right\}_{n=1}^\infty$ is countably dense in $L^1$. Hence,
$$
\psi^\pm_{0,A,B}(k,x)\, \phi(k)=0, \forall k \in (0,\infty)\setminus O. 
$$
Finally this implies that $ \|\left(F^\pm_{0,A,B}\right)^\star\, \phi\|=\|\phi\|=0$.

\bull

Let us define,

\beq \label{12.33}
\psi^\pm_{A,B}(k,x):=\psi^\pm_{0,A,B}(k,x)- R_{A,B}(k^2 \mp i0)\, V \,\psi^\pm_{0,A,B}(k,x).
\ene

\begin{lemma} Suppose that $ V$ satisfies \eqref{1.6}, \eqref{1.7}. Then, with $\psi_{A,B}$ defined in \eqref{3.13},
\beq \label{12.33b}
 \psi^\pm_{A,B}(k,x)= \psi_{A,B}(\mp k, x).
 \ene
\end{lemma} \noindent{\it Proof:} By \eqref{12.25},  \eqref{12.28},
\beq \label{12.34}
\left(F_{A,B}^\pm \, \psi \right)(k)= \, \sqrt{\frac{2}{\pi}}\, \int_0^\infty\,   \left( \psi^\pm_{A, B}(k,x)\right)^\dagger \, \psi(x)\, dx, \qquad \psi \in L^2 \cap L^1,
\ene

\beq \label{12.35}
\left( \left(F_{A,B}^\pm \right)^\dagger \psi \right)(x)= \, \sqrt{\frac{2}{\pi}}\, \int_0^\infty\ \, \psi^\pm_{A, B}(k,x) \, \psi(k)\, dk, \qquad \psi \in L^2 \cap L^1.
\ene
Furthermore, by \eqref{11.11}, \eqref{12.5b}, \eqref{12.5c},  \eqref{12.15}\eqref{12.22}, \eqref{12.33} 

\beq\label{12.36} 
\psi^\pm_{A,B}(k,x)= \frac{1}{2} \, e^{\pm ikx}+  e^{\mp i kx} \,D_{\pm}(k) +o(1), \qquad \textrm{as}\, x \rightarrow \infty,
\ene
for a matrix $D(k)$. 
Let us define,
$$
\phi^\pm(k,x):=  \frac{1}{2} \, f(\pm k,x)+   f(\mp k,x) D_{\pm}(k).
$$

Then, $\psi^\pm_{A,B}(k,x)-  \phi^{\pm}(k,x) $ is a solution to \eqref{3.1} and by \eqref{3.2}, \eqref{12.36},   it  goes to zero as $ x \rightarrow \infty$.  Moreover,  it is a linear combination of $f(k,x)$ and $g(k,x)$ (recall that the combined $2n$ columns of $f(k,x)$ and of $g(k,x)$ form a fundamental set of solutions to  \eqref{3.1} ). As  $f( k, x)$ behaves as $ e^{ i kx}$ and $g( k, x)$ behaves as $ e^{- i kx}$, as $ x \rightarrow \infty$,  necessarily, $\psi^\pm_{A,B}(k,x)-  \phi^\pm(k,x)=0 $, and then,
\beq\label{12.37}
\psi^\pm_{A,B}(k,x)= \frac{1}{2} \, f(\pm k,x)+\, f(\mp k,x)\, D_{\pm}(k).
\ene

Hence, by \eqref{3.16} $ \chi^\pm (k,x):= \psi^\pm_{A,B}(k,x)- \psi_{A,B}(\mp k,x)= f(\mp k, x)(D_{\pm}(k)- \frac{1}{2} S(\mp k))$. As $\chi^\pm(k,x)$ satisfies the boundary condition \eqref{1.2}, by \eqref{3.11},
$$
-B^\dagger \, \chi^{\pm} (k,0)+ A^{\dagger} \chi{^\pm}'(k,0)= - J_{A,B}^\dagger(\pm k)(D_{\pm}(k)- \frac{1}{2}S(\mp k))=0.
$$ 
But as $J^\dagger_{A,B}(k)$ is invertible for $k\in \mathbb R \setminus 0$, $D_{\pm}(k)= \frac{1}{2}\, S(\mp k)$, and then, by \eqref{3.16}, \eqref{12.37},

\beq \label{12.39}
 \psi^\pm_{A,B}(k,x)= \psi_{A,B}(\mp k, x).
 \ene

\section{The Wave operators}\sss

We take as unperturbed operator $ H_0:=  H_{-I,0}$ , with  the Neumann boundary condition,
$\varphi'(0)=0$. If we interpret the matrix Schr\"odinger operator as a star graph with one vertex the Neumann boundary condition corresponds physically to the case where there is no transfer of current  between the different wires of the graph, what is a natural unperturbed boundary condition. We denote by $F_0$ the generalized Fourier map of $H_0$, i.e., $ F_0= F^\pm_{0, -I,0}$. Note that  $ F^+_{0,-I,0}=F^-_{0, -I,0}$ and that $F_0$ is just the cosine transform,
\beq\label{13.1}
\left(F_{0} \psi \right)(k):= \sqrt{\frac{2}{\pi}}\, \int_0^\infty\, \cos(kx)\, \psi(x)\, dx, \qquad \psi \in L^2.
\ene
Furthermore,  $H_0$ has no eigenvalues.

\subsection{The case of zero potential}
\begin{theorem}\label{theo13.1}
The wave operators $W_\pm(H_{0,A,B}, H_0 )$ exist and are complete, they are  isometric with initial subspace $L^2$ and final subspace $ \mathcal H_{\textrm{ac}}(H_{0,A,B})$   . Furthermore,

\beq\label{13.2}
W_{\pm}(H_{0,A,B}, H_0)= \left(F^\pm_{0,A,B}\right)^\dagger\, F_0.
\ene
\end{theorem}

\noindent{\it Proof:} Once formula \eqref{13.2} is proven, the fact that $W_\pm(H_{0,A,B}, H_0)$ 
 are  isometric with initial subspace $L^2$ and final subspace $ \mathcal H_{\textrm{ac}}(H_{0,A,B})$ follows from Theorem \ref{theo12.2}. By \eqref{12.20}, \eqref{12.29} it is enough to prove that,

\beq \label{13.3} 
\left\| \left[\left(F^\pm_{0,A,B}\right)^\dagger - \
F_0^\dagger \right]\, e^{-it k^2} \,\phi(k) \right\| \rightarrow 0, \, \textrm{as}\, t \rightarrow \pm \infty, \qquad \forall \phi \in C^\infty_0((0, \infty)). 
\ene
But, by  by \eqref{12.5b}, \eqref{12.5c},\eqref{12.9} \eqref{12.15}, \eqref{12.16}  and \eqref{13.1} this is equivalente to,

\beq\label{13.4}
 \left\|   \int_0^\infty\, e^{\mp ik x}\, e^{-it k^2}\, \phi(k)\, dk\right\| \rightarrow 0,   \, \textrm{as}\, t \rightarrow \pm \infty, \qquad \forall \phi \in C^\infty_0((0, \infty)). 
\ene
Equation \eqref{13.4} holds by Lemma 2.6.4  of \cite{ya1}.

\subsection{The relative wave operators}
We consider now the relative wave operators $W_\pm(H_{A,B}, H_{0,A,B})$ that will allow us to obtain  the wave operators with integrable potential by the chain rule,  $W_\pm (H_{A,B}, H_0)= W_\pm(H_{A,B}, H_{0,A,B})\, W_{\pm}(H_{0,A,B}, H_0)$.

\begin{theorem} \label{theo13.2} Suppose that $ V$ satisfies \eqref{1.6}, \eqref{1.7}.
Then, the wave operators $ W_\pm(H_{A,B}, H_{0,A,B})$ exist and are complete, they are partially isometric with initial subspace $ \mathcal H_ {\textrm{ac}}(H_{0,A,B})$ and final subspace $\mathcal H_{\textrm{ac}}(H_{A,B})$. Furthermore,
\beq \label{13.5}
W_\pm(H_{A,B}, H_{0,A,B})=  \left(F^\pm_{A,B}\right)^\dagger\, F^\pm_{0, A, B}.
\ene
\end{theorem}

\noindent{\it Proof:} Once \eqref{13.5} is proven, the rest of the theorem follows from Theorems \ref{theo12.2} and \ref{theo12.3}. We prove \eqref{13.5} for $W_+(H_{A,B}, H_{0,A,B})$. The case  $W_-(H_{A,B}, H_{0,A,B})$  similar. Since $\left\| e^{it H_{A,B}} \, e^{-it H_{0,A,B}}\, \phi  \right\|=  \left\|   \left(F^\pm_{A,B}\right)^\dagger\, F^\pm_{0, A, B}\, \phi  \right\|= \left\|  \phi \right\|$, for all $ \phi \in \mathcal H_{ \textrm{ac}}(H_{0,A,B})$ it is enough to prove that
\beq\label{13.6}
\lim_{t \rightarrow \infty}\, \left( e^{-it H_{A,B}}\, E_{A,B}( O_0) \psi_0, e^{-it H_{0,A,B}}\, E_{0,A,B}(O_1)\, \psi_1\right)=
 \left( F^+_{A,B}\, E_{A,B}( O_0) \psi_0, \, F^+_{0,A,B}\,  E_{0,A,B}(O_1)\, \psi_1  \, \right),
\ene
for all $ O_0=(k^2_0, k^2_1), O_1= (k^2_2, k_3^2), 0 < k_0 < k_1, 0 < k_2 < k_3$ and all $ \psi_0, \psi_1 \in L^2 \cap L^1.$

By \eqref{11.22}, \eqref{12.22},  for $ \lambda >0$,
\beq \label{13.7}
R_{A,B}(\lambda+ i\varepsilon)- R_{A,B}(\lambda-i \varepsilon)= \left[  R_{0,A,B}(\lambda+ i\varepsilon)- R_{0,A,B}(\lambda-i \varepsilon) \right]\, L^+(\lambda+i \varepsilon)+  R_{0,A,B}(\lambda-i \varepsilon)\, V_1\, J(\lambda+i \varepsilon),
\ene
where, $L^+(\lambda+i \varepsilon)$ is defined in \eqref{12.21} and
\beq \label{13.8}
J(\lambda+i\varepsilon ):=\, - V_2 \, \left[ R_{A,B}(\lambda+ i\varepsilon)- R_{A,B}(\lambda-i \varepsilon) \right].
\ene
Then, by Stone's formula,
\beq \label{13.9}
\frac{d}{d\, \lambda}\,\left(E_{A,B}(\lambda)\, \psi_0, e^{-it H_{0,A,B}}\, E_{0,A,B}(O_1)\, \psi_1\right)= \lim_{\varepsilon \rightarrow 0}\, T_1(\varepsilon)+ \lim_{\varepsilon \rightarrow 0}\, T_2(\varepsilon),
\ene
where,
\beq \label{13.10}
T_1(\varepsilon):= \left(L^+(\lambda+ i\varepsilon)\,  \psi_0,\,   \frac{1}{2 \pi i} \,\left[  R_{0,A,B}(\lambda+ i0)- R_{0,A,B}(\lambda-i 0) \right]\, e^{-it H_{0,A,B}}\, E_{0,A,B}( O_1)\, \psi_1\right), 
\ene
\beq \label{13.11}
T_2(\varepsilon) := \frac{1}{2 \pi i}\, \left( J(\lambda+i \varepsilon)\, \psi_0, \int_{O_1}\, \frac{e^{-it \rho}}{\rho-(\lambda+i \varepsilon)}\,  \frac{1}{2 \pi i}\, V_1\,\left[  R_{0,A,B}(\rho+ i0)- R_{0,A,B}(\rho-i 0) \right]\, \psi_1\right)\, d \rho,
\ene
where we used Theorem \ref{theo11.1}.  By \eqref{12.16}, \eqref{12.18}, and \eqref{12.26}  we have that,
\begin{eqnarray} \label{13.12}
\lim_{\varepsilon \rightarrow 0}\, T_1(\varepsilon)&=& e^{it \lambda}\, \chi_{O_1}(\lambda)\, \left( L^+(\lambda)\, \psi_0,
\frac{1}{2 \pi i}\, \left[  R_{0,A,B}(\lambda+ i0)- R_{0,A,B}(\lambda-i 0) \right]\, \psi_1 \right)  \nonumber\\
&=& \frac{1}{2 \sqrt{\lambda}}\,  e^{it \lambda}\, \chi_{O_1}(\lambda)\,   \left< \left(F^+_{A,B}\, \psi_0 \right)(\sqrt{\lambda}).  \left(F^+_{0,A,B}\, \psi_1\right)(\sqrt{\lambda})\,\right>.
\end{eqnarray}
Let us denote,
\beq \label{13.13}
I_t ( \lambda ,\varepsilon):= \int_{O_1}\, \frac{ e^{-it \rho}}{\rho-(\lambda+i \varepsilon)}\,G(\rho), 
\ene
where,
$$
G(\rho):=\, \frac{1}{2 \pi i} \,V_1 \left[  R_{0,A,B}(\rho+ i0)- R_{0,A,B}(\rho-i 0) \right]\, \psi_1  \in L^2.
$$

Then, denoting by $F$ the  Fourier transform   in $L^2\left( \mathbb R, L^2\right)$ , it follows  that,
\beq\label{13.14}
I_t(\lambda):= \lim_{\varepsilon \rightarrow 0} \, I_t(\lambda, \varepsilon) = F^{-1}\,\left(2\pi i\,\chi_{(0,\infty)}(k)  \, \left[\displaystyle F\, \chi_{O_1}(\rho)\, e^{-it \rho}\, G(\rho) \right](k) \right)(\lambda),
\ene   
where the limit exists in $L^2(\mathbb R, L^2)$.
Hence, by \eqref{13.9}, \eqref{13.12}, \eqref{13.13} and \eqref{13.14}
\beq
\nonumber  \left( e^{-it H_{A,B}}\, E_{A,B}( O_0) \psi_0, e^{-it H_{0,A,B}}\, E_{0,A,B}(O_1)\, \psi_1\right)= \int_{O_0}\, e^{-it\lambda}\, \frac{d}{d \lambda}\,\left( E_{A,B}(\lambda)\, \psi_0, e^{-it H_{0,A,B}}\, E_{0,A,B}(O_1)\, \psi_1\right)\, d \lambda=
\ene
\beq\nonumber
 \int \chi_{O_0 \cap O_1}(k^2)    \left< \left(F^+_{A,B}\, \psi_0 \right)(k),  \left(F^+_{0,A,B}\, \psi_1\right)(k)\,\right> \, dk +\frac{1}{2 \pi i}\, \int_{O_0} \left( J(\lambda+i 0)\, \psi_0, I_t(\lambda)\right)=
 \ene
 \beq\label{13.15}
 \left(F^+_{A,B}\,E_{A,B}(O_0)\, \psi_0 ,\,F^+_{0,A,B}\, E_{A,B}(O_1)\,\psi_1\,\right) +\frac{1}{2 \pi i}\, \int_{O_0} \left( J(\lambda+i 0)\, \psi_0, I_t(\lambda)\right).
 \ene
 Furthermore,
 $$
 \left\| I_t(\rho) \right\|^2_{L^2\left(\mathbb R, L^2\right)}=  4 \,\pi^2 \,\int_t^\infty\,  \left\|\left( F \left[\chi_{O_1}(\rho)\,  G(\rho)\, \right]\right)(k)
 \right\|_{L^2}^2 \, dk    \rightarrow 0 \,\textrm{as}\, t\rightarrow \infty,
 $$
 and then,
 \beq\label{13.16}
 \lim_{t \rightarrow \infty} \, \int_{O_0} \left( J(\lambda+i 0)\, \psi_0, I_t(\lambda)\right)=0.
\ene
By \eqref{13.15} and \eqref{13.16} equation  \eqref{13.6} holds.

\subsection{The wave operators}

\begin{theorem} \label{theo13.3} Suppose that $ V$ satisfies \eqref{1.6}, \eqref{1.7}.
Then, the wave operators $ W_\pm(H_{A,B}, H_0)$ exist and are complete, they are  isometric with initial subspace $L^2 $ and final subspace $\mathcal H_{\textrm{ac}}(H_{A,B})$. Furthermore,
\beq \label{13.17}
W_\pm(H_{A,B}, H_{0})=  \left(F^\pm_{A,B}\right)^\dagger\, F_{0}.
\ene
\end{theorem}

\noindent{\it Proof:} By the chain rule,
$$
W_\pm(H_{A,B}, H_0)= W_\pm(H_{A,B}, H_{0,A,B})\, W_\pm(H_{0,A,B}, H_0).
$$
Then, the theorem follows from Theorems \ref{theo13.1} and \ref{theo13.2}.

\bull
\section{The scattering operator and the scattering matrix}\sss 
 
 The scattering operator is defined as,
 \beq \label{14.1}
 \mathcal{S}_{A,B}:= \left(W_+(H_{A,B}, H_0)\right)^\dagger\, W_-(H_{A,B}, H_0).
 \ene
 By theorem \ref{theo13.3} $S$ is unitary in $L^2$ and,
 \beq \label{14.2}
 \mathcal{S}_{A,B}= F_0^\dagger\, F^+_{A,B}\, \left(F^-_{A,B}\right)^\dagger\, F_{0}.
\ene
We denote,
\beq\label{14.3}
\hat{ \mathcal S}_{A,B}:= F_0\, \mathcal{S}_{A,B}\, F_0^\dagger= \, F^+_{A,B}\, \left(F^-_{A,B}\right)^\dagger.
\ene
Since $\hat{\mathcal S}_{A,B}$ commutes with $H_0$ it decomposes as a direct integral in the spectral representation of $H_0$.
Then,  for $ k \in (0,\infty)$, there is a unitary $ n \times n$  matrix $\hat{\mathcal {S}}_{A, B}(k)$, such that,
\beq\label{14.4}
\left( \hat{\mathcal S}_{A,B} \psi\right)(k)= \hat{\mathcal{S}}_{A,B}(k)\, \psi(k), \quad \forall \psi \in L^2.
\ene
Recall that the scattering matrix $S_{A,B}(k)$ was defined in \eqref{3.14}.
\begin{theorem}\label{theo14.1} Suppose that $ V$ satisfies \eqref{1.6}, \eqref{1.7}. Then,
\beq \label{14.5}
\hat{\mathcal{S}}_{A,B}(k)= S_{A,B}(k), k \in (0,\infty).
\ene
\end{theorem}
 
 \noindent{\it Proof:}  By \eqref{12.34}, \eqref{12.35}, \eqref{12.39}, \eqref{14.3} and \eqref{14.4}
 $$
\hat{ \mathcal S}_{A,B}(k)\, \left( \psi^{-}_{A,B}(k,x)\right)^\dagger=  \left(\psi^{+}_{A,B}(k,x)\right)^\dagger,
 $$
 and then,
 \beq \label{14.6}
   \psi^{-}_{A,B}(k,x)\, \hat{ \mathcal S}^\dagger_{A,B}(k)\,=  \psi^{+}_{A,B}(k,x).
   \ene
   Hence, by \eqref{3.15}, \eqref{12.39},
   $$
f(-k,x)\, \left[\hat{\mathcal S}_{A,B}^{\dagger}(k)- S_{A,B}^{\dagger}(k)\right]= - f(k,x)\, \left[ S_{A,B}(k)\, \hat{\mathcal S}_{A,B}^{\dagger}(k)-I\right].
$$
but, as $f(\pm k,x)\approx e^{\pm i kx}, x \rightarrow \infty$, we have that, $\hat{\mathcal S}_{A,B}(k)=  S_{A,B}(k)$.

\bull

 By \eqref{3.7}
 $$
 \varphi_{A,B}(k,x)= f(k,x)\,\alpha+ g(k,x)\beta.
 $$
 Then, by \eqref{3.3}, \eqref{3.4} \eqref{3.10},
 $$
 \alpha= \frac{1}{2ik}\, \left[  g(-k^\star,x)^\dagger; \varphi_{A,B}(k,x)\right], \qquad \beta= -\frac{1}{2ik}\, J_{A,B}(k).
 $$
 It follows that, for $\textrm{Im}\, k >0$,
 \beq \label{14.7}
 \lim_{x \rightarrow \infty}\, e^{ikx}\, \left[  \varphi_{A,B}'(k,x)-i k \varphi_{A,B}(k,x) \right]= J_{A,B}(k).
 \ene  
  Then, by \eqref{3.1}, \eqref{3.8}, \eqref{3.27} and \eqref{14.7}, for $ \textrm{Im}\, k>0$,
    
  \begin{equation}\label{14.8}\begin{array}{l}
  \int_0^\infty\, e^{ikx} \, V(x)\, \varphi_{A,B}(k,x)\,dx  = \lim_{R \rightarrow \infty}\, \int_0^R\, e^{ikx}\, V(x)\, \varphi_{A,B}(k,x)\, dx=\\\\
 \lim_{R \to \infty}  \int_0^R\, e^{ikx}\, \left( \varphi_{A,B}''(k,x)\,+ k^2\,\varphi_{A,B}(k,x)\right)\,dx = J_{A,B}(k)- J_{0,A,B}(k),
   \end{array}\ene

   where in \eqref{14.8} we integrated twice by parts. Then, by continuity in $k$
   \beq\label{14.9}
  J_{A,B}(k)= J_{0,A,B}(k)+ \int_0^\infty\, e^{ikx}\, V(x)\, \varphi_{A,B}(k,x)\, dx, k \in \overline{\mathbf C^+}.
  \ene

  \begin{prop}\label{prop14.2} Suppose that $ V$ satisfies \eqref{1.6}, \eqref{1.7}. Then,
  \beq \label{14.10}
  S_{A,B}(k)= S_{0,A,B}(k) -\frac{ 2i} {k} \int_0^\infty\, \psi_{0, A,B}\, V(x)\, \psi_{A,B}(k,x)\,dx,  k \in \mathbb R\setminus 0.
  \ene
  \end{prop}
  
   \noindent{\it Proof} By \eqref{3.13},  \eqref{3.14}, \eqref{3.16} with $V=0$, \eqref{3.27} and \eqref{14.9}
   
   \beq\label{14.11}
   \int_0^\infty\, \psi_{0, A,B}\, V(x)\, \psi_{A,B}(k,x)\,dx= -\frac{ik}{2}\, \left(- S_{A,B}(k)+ S_{0,A,B}(k)\right).
   \ene
   Equation \eqref{14.10} follows from \eqref{14.11}.

    \section{The spectral shift function}\sss
  Lets us denote by $R_0(z)$ the resolvent of $H_0$, i.e. $R_0(z):= \left( H_0 -z \right)^{-1}, z \in \rho(H_0).$
  
  \begin{lemma} \label{lemm15.1} Suppose that $ V$ satisfies \eqref{1.6}, \eqref{1.7}. Then, for every $ z \in \rho\left(H_{A,B}\right) \cap \rho\left(H_{0,A,B}\right)$ the difference of resolvents, $ R_{A,B}(z)- R_{0,A,B}(z)$ is trace class and for evey   $ z \in \rho\left(H_{0,A,B}\right) \cap \rho\left(H_{0}\right)$,  $ R_{0,A,B}(z)- R_{0}(z)$ is trace class.
  \end{lemma}
  
  \noindent{\it Proof:} By \eqref{11.6b}  $V_2\, R_{0,A,B}(z)$ is Hilbert-Schmidt.  Since $D(H_{A,B})$  is contained in $\mathbf{H}_1$ and the imbedding of $\mathbf{H}_1$ into $L^\infty$ is bounded (see \eqref{10.7b}), 
  
  $$
  \left|  \left(R_{A,B}(z) \varphi \right)(x)\right| \leq C \| \varphi\|_{L^2}, \qquad \forall x \in \mathbb R^+.
  $$
  Then, by the Riez representation theorem, for all $ x \in \mathbb R^+$ there is a $\psi_x \in L^2$ such that,
  $$
   \left(R_{A,B}(z) \varphi \right)(x)= \left(\varphi, \psi_x \right),
  $$
  and 
  $$
  \left\| \psi_x  \right\|_{L^2} \leq C, \qquad \forall x \in \mathbb R^+. 
  $$
  Hence, $ V_1\, R_{A,B}(z) $ is an integral operator with the Hilbert-Schmidt kernel,
  $$
  V_1(x)\, \psi_x(y), \qquad x,y \in \mathbb R^+.
  $$
   
Hence, $ V_1\, R_{A,B}(z) $ is Hilbert-Schmidt. Furthermore as $ V_1\, R_{A,B}(z)$ and    $V_2\, R_{0,A,B}(z)$ are Hilbert-Schmidt, by  \eqref{11.16}
  $ R_{A,B}(z)- R_{0,A,B}(z)$ is trace class.
  
  Furthermore using the identity,
  $$
  \left( -\frac{d^2}{d x^2} \varphi , \psi \right)- \left( \varphi , -\frac{d^2}{d x^2} \psi\right)= \varphi'(0)\, \overline{\psi}(0)- 
 \varphi(0) \overline{\psi'}(0), \qquad \varphi, \psi, \in \mathbf{H}_2,
  $$
  with $\varphi= R_{0}(z)\, f, \psi= R_{0,A,B}(\overline{z})\,g,  f, g \in L^2$ we prove that
  $$
  R_{0,A,B}(z)- R_{0}(z)= \left( T R_{0,A,B}(\overline{z}) \right)^\dagger\, \left( T \frac{d}{dx}\, R_{0}(z)\right)- 
   \left( T\, \frac{d}{d x}  R_{0,A,B}(\overline{z}) \right)^\dagger\, \left( T \, R_{0}(z)\right),
  $$
  where $T$ is the bounded trace operator from $\mathbf{H}_1$ into $\mathbf C^n$,
  $$
  T \varphi = \varphi(0).
  $$
  Furthermore, since $H_0$  has Neumann boundary condition, $T \frac{d}{dx}\, R_{0}(z)=0$. Hence,
  $$
   R_{0,A,B}(z)- R_{0}(z)= -  \left( T\, \frac{d}{d x}  R_{0,A,B}(\overline{z}) \right)^\dagger\, \left( T \, R_{0}(z)\right).
  $$

  Let us denote by $K(z) :=T\, R_{0}(z), N(\overline{z}):= T\, \frac{d}{dx} R_{0,A,B}(\overline{z})$. Then by \eqref{11.7},    $K(z)$ and $N(\overline{z})$ are Hilbert-Schmidt and
  $$
  \left\| K(z) \right\|_2 \leq C D(\sqrt{z}) \, \frac{1}{\sqrt{\textrm{Im}\, \sqrt{z}}}, \qquad \left\| N\left(\overline{z}\right) \right\|_2 \leq C  F(\sqrt{\overline{z}})  \frac{1}{ \sqrt{\textrm{Im}\,\sqrt{\overline{z}}}}.
  $$
  Hence,  $R_{0,A,B}(z)- R_{0}(z)$ is trace class and
  \beq \label{15.1}
  \left\|    R_{0,A,B}(z)- R_{0}(z)  \right\|_1 \leq C D(\sqrt{z})\, F(\sqrt{\overline{z}}) \, \frac{1}{{\textrm Im}\, \sqrt{z}}.
  \ene
   
  \bull
 
 For later use, we note that for all $\alpha > 1/4$
 \beq\label{15.2}
  \left\|  R_{A,B}(-E)- R_{0,A,B}(-E)  \right\|_1 \leq C \, E^{-2+ 2 \alpha},  \,\textrm{as}\, E \rightarrow \infty.
 \ene
 This estimate is proven  as in the proof of Lemma 5.6 in page 194 of \cite{ya2},  observing that $ |V|^{1/2}\, (H_0+I)^{-\alpha}$ is Hilbert-Schmidt for $ \alpha > 1/4$. Indeed, with $F_0$ as in \eqref{13.1},  $|V|^{1/2}\, (H_0+I)^{-\alpha}= |V|^{1/2}\,F_0^\dagger (k^2+1)^{-\alpha}\, F_0$, and the operator   $|V|^{1/2}\, F_0^\dagger (k^2+1)^{-\alpha} $ is an integral operator with the Hilbert-Schmidt kernel  $\sqrt{\frac{2}{\pi}}, |V(x)|^{1/2}\, \cos(kx) \,  (k^2+1)^{-\alpha}$.

  For any pair of self-adjoint operators $  F,G$ we denote their spectral shift function by $\xi(E; F, G)$.
 
 As the difference of the resolvents, respectively, of $H_{A,B}$ and $H_0$,   $H_{0,A,B}$ and $H_0$,   $H_{A,B}$ and $H_{0,A,B}$,  are trace class we can use the abstract theory of section 9 of chapter 0 of \cite{ya2} to define their spectral shift functions  as in  equation (9.26), page 50 of \cite{ya2}. Take $E_0 > $ such that,  $H_{A,B}+ E_0 >I, H_{0,A,B} + E_0 >I$. Since $H_0 \geq 0$, we also have that $H_0 +E_0 >I$.Then,

 \begin{eqnarray} \label{15.3}
 \xi(E; H_{A,B}, H_0)  &=& - \xi\left( (E+ E_0)^{-1};  \left( H_{A,B}+ E_0\right)^{-1},   \left( H_{0}+ E_0\right)^{-1}   \right), \, E \geq -E_0, \\
  \xi(E; H_{A,B}, H_0)&=&0, \, E < -E_0,\label{15.4}
  \end{eqnarray}
    \begin{eqnarray} \label{15.5}
 \xi(E; H_{0,A,B}, H_0)  &=& - \xi\left( (E+ E_0)^{-1};  \left( H_{0,A,B}+ E_0\right)^{-1},   \left( H_{0}+ E_0\right)^{-1}   \right), \, E \geq- E_0, \\
  \xi(E; H_{0,A,B}, H_0)&=&0, \, E < -E_0,\label{15.5c}
  \end{eqnarray}
  
   \begin{eqnarray} \label{15.5b}
 \xi(E; H_{A,B}, H_{0,A,B})  &=& - \xi\left( (E+ E_0)^{-1};  \left( H_{A,B}+ E_0\right)^{-1},   \left( H_{0,A,B}+ E_0\right)^{-1}   \right), \, E \geq- E_0, \\
  \xi(E; H_{A,B}, H_{0,A,B})&=&0, \, E < -E_0. \label{15.6}
  \end{eqnarray}

 By the addition formula for the spectral shift function (see Proposition 5 in page 275 of \cite{ya1} ),
 \beq \label{15.7}
   \xi(E; H_{A,B}, H_0)=  \xi(E; H_{A,B}, H_{0,A,B}) +  \xi(E; H_{0,A,B}, H_0).
   \ene
   Furthermore, we have that,
   \beq \label{15.8}
   \int_{-\infty}^\infty\, \left| \xi(E; H_{A,B}, H_{0,A,B})\right|\, (1+|E|)^{-1/2-\varepsilon}\, dE  < \infty , \qquad \forall \varepsilon >0.
   \ene 
   Equation \eqref{15.8} is proven, using \eqref{15.2},  Theorem 9.7 in page 50 of \cite{ya2} and Remark 9.9 in page 51 of
   \cite{ya2}.

  To compute  $ \xi(E; H_{0,A,B}, H_0)$ we can use the diagonal representation with the matrices $\tilde{A}, \tilde{B}$ in \eqref{3.25}. Let us denote, respectively, by $-\Delta_{\textrm D}, -\Delta_{\textrm N}, -\Delta_\gamma$ the selfadjoint realizations of $-\frac{d^2}{d x^2}$ in $L^2((0,\infty), \mathbf{C}) $ with Dirichlet boundary condition, $\varphi(0)=0$, Neumann boundary condition, $\varphi'(0)=0$ and with mixed boundary condition, $\varphi'(0)= \gamma\,\varphi(0)$.   
  
  With the notation of \eqref{3.25b} we designate,
  \beq \label{15.9}
  \gamma_j= -\cot \theta_j, \forall \theta_j \neq \pi, \pi/2.
  \ene
  Then,
  \beq\label{15.10}
  \xi(E; H_{0,A,B}, H_0)= n_{\textrm D}\, \xi(E;-\Delta_{\textrm D}, -\Delta_{\textrm N})+ \tilde{\sum}_{j=1}^n \xi(E; -\Delta_{\gamma_j}, -\Delta_{\textrm N}),
  \ene
  where by $\tilde{\sum}_{j=1}^n$ we denote the sum over all $1 \leq j \leq n$ with $ \theta_j \neq \pi, \pi/2$. Note that $\xi(E; -\Delta_{\textrm N}, - \Delta_{\textrm N})=0$. The spectral shift functions,  $\xi(E;-\Delta_{\textrm D}, -\Delta_{\textrm N}),  \xi(E; -\Delta_{\gamma}, -\Delta_{\textrm N})$ can be easily computed as in Section 7 of Chapter 4 of \cite{ya2} where $   \xi(E; -\Delta_{\gamma}, -\Delta_{\textrm D})$ is calculated. Observe that,   $\xi(E; -\Delta_{\textrm D}, -\Delta_{\textrm N}) = - \xi(E; -\Delta_{\textrm N}, -\Delta_{\textrm D})$. The result is as follows,
  
  \begin{eqnarray}
  \label{15.11}
  \xi(E; -\Delta_{\textrm D}, -\Delta_{\textrm N})&=& \frac{1}{2},\\
  \label{15.12}
  \xi(E; -\Delta_{\gamma}, -\Delta_{\textrm N})&=& \frac{1}{2} - \frac{1}{\pi}\, \arctan\left( \frac{\sqrt{E}}{\gamma}  \right), \qquad \textrm{if}\, \gamma >0,
   \\
   \label{15.13}
   \xi(E; -\Delta_{\gamma }, -\Delta_{\textrm N})&=& - \frac{1}{2} - \frac{1}{\pi}\, \arctan\left( \frac{\sqrt{E}}{\gamma}  \right), \qquad \textrm{if}\, \gamma < 0.
   \end{eqnarray}
   Actually,   \eqref{15.12}, \eqref{15.13}  follow from the computation in Section 7 of Chapter 4 of \cite{ya2} since by the addition formula (Proposition 5 in page 275 of \cite{ya1}),
   $ \xi(E; -\Delta_{\gamma}, -\Delta_{\textrm N})=  \xi(E; -\Delta_{\gamma}, -\Delta_{\textrm D})+  \xi(E; -\Delta_{D}, -\Delta_{\textrm N})$.

By  \eqref{15.10}, \eqref{15.11}, \eqref{15.12},\eqref{15.13} it follows that,
\beq \label{15.14}
\lim_{E \rightarrow \infty} \, \xi(E; H_{0,A,B}, H_0)=  = \frac{1}{2}\,  n_{\textrm D}.
\ene

 By \eqref{3.33},\eqref{3.34},\eqref{3.36}
\beq\label{15.15}
\textrm{det}\, S_{A,B}(k)= \textrm{det}S_{A,B}(\infty)+ O\left( 1/|k|\right) \, \textrm{as}\,  k \rightarrow \infty, 
\ene
where
\beq\label{15.16}
\textrm{det} S_{A,B}(\infty)= (-1 )^{n_{\textrm D}}.
\ene

Then,
\beq\label{15.17}
\lim_{k \rightarrow \infty}\, \textrm{arg}\left[\textrm{det}\, S_{A,B}(k)  \right]=\begin{cases} 0+2\pi j, j=0, \pm 1,\pm2,\cdots,\qquad \textrm{if}\, n_{\textrm D}\, \textrm{is even},\\\\
\pi+ 2 \pi j,  j=0, \pm 1,\pm2,\cdots, \qquad \textrm{if}\, n_{\textrm D}\, \textrm{is odd}.
\end{cases}
\ene

By the Birman-Krein formula (see equation (9.10) in page 47 of \cite{ya2})
\beq\label{15.18}
\textrm{det}\, S(\sqrt{E})= e^{-2\pi i\, \xi(E; H_{A,B}, H_0)}, \quad E >0.
\ene  
By \eqref{15.18} $, -2 \pi\,  \xi(E; H_{A,B}, H_0)$ coincides, for $ E >0$, with one of the continuous branches of $\textrm{det}\, S(\sqrt{E})$.  By \eqref{15.17} this implies  that $\lim_{k \rightarrow \infty}\,   \xi(E; H_{A,B}, H_0) $ exits, and by  \eqref{15.7}, \eqref{15.14}, also  $\lim_{k \rightarrow \infty}\,   \xi(E; H_{A,B}, H_{0,A,B}) $ exits, but then, by \eqref{15.8},
  
  \beq \label{15.19}
  \lim_{k \rightarrow \infty}\,   \xi(E; H_{A,B}, H_{0,A,B})=0.
\ene
Hence, by \eqref{15.7}, \eqref{15.14}. \eqref{15.19}
\beq \label{15.20} 
 \lim_{k \rightarrow \infty}\,   \xi(E; H_{A,B}, H_{0})=   \frac{1}{2}\,n_{\textrm D}.
 \ene
.
Let us denote by $\Theta (k)$ the continuous branch of $\textrm{arg}\left[\textrm{det}\, S(k)  \right]$ such that,

\beq \label{15.21}
\lim_{k \rightarrow \infty}\, \Theta(k)= - \pi n_{\textrm D}.
\ene
If $ n_{\textrm D}$ is even this corresponds to $j= -  n_{\textrm D}/2$ in \eqref{15.17} and to $j= -  (n_{\textrm D}+1)/2$ if
 $ n_{\textrm D}$ is odd . 
Then, by \eqref{15.18}, \eqref{15.20} and \eqref{15.21} we have that,

\beq\label{15.22}
 \xi(E; H_{A,B}, H_0)= - \frac{1}{2 \pi}\, \Theta(\sqrt{E}) \,, \quad E >0.
\ene  
By \eqref{15.15},\eqref{15.20},\eqref{15.21} and \eqref{15.22},
\beq\label{15.23}
 \xi(E; H_{A,B}, H_0) = \frac{n_{\textrm D}}{2}+ O\left(\frac{1}{\sqrt{E}}  \right), E \rightarrow \infty.
\ene

By \eqref{15.23}
\beq \label{15.24}
\begin{cases}
   \int_{-\infty}^\infty\, \left| \xi(E; H_{A,B}, H_{0})\right|\, (1+|E|)^{-1-\varepsilon}\, dE  < \infty, \qquad \forall \varepsilon >0, \,\textrm{if} \, n_{\textrm D} >0,\\ \\
   
     \int_{-\infty}^\infty\, \left| \xi(E; H_{A,B}, H_{0})\right|\, (1+|E|)^{-1/2-\varepsilon}\, dE < \infty, \qquad \forall \varepsilon >0, \, \textrm{if} \, n_{\textrm D} =0.
     \end{cases}
   \ene 

Let us denote by $\{ E_j \}_{j=1}^ P$ the eigenvalues of $H_{A,B}$  in increasing order
$ E_1 < E_2 < \cdots <E_j < \cdots$, with $P$ finite or infinite. Let  $m_j$ be the multiplicity of $E_j$. Then by Proposition 9.2 in page 46 of \cite{ya2} for $ E <0$, $ \xi(E;H_{A,B}, H_0)$ is piecewise constant and it assumes  integral values. Furthermore, 

\begin{eqnarray}\label{15.25}
 \xi(E;H_{A,B}, H_0)&=&0,\, \textrm{for}\, E < E_1,\\ \label{15.26}
   \xi(E;H_{A,B}, H_0)&=&-m_1,\, \textrm{for}\,  E_1<  E < E_2, \cdots,  \xi(E;H_{A,B}, H_0)=
   - \sum_{j=1}^l\, m_j  \, \textrm{for}\,  E_l<  E < E_{l+1}.
   \end{eqnarray}
  Furthermore, by \eqref{15.24}, \eqref{15.25} and \eqref{15.26}
  \beq \label{15.27}
  \sum_{j=1}^{P-1}\, m_j\, (E_{j+1}- E_j)  <  \infty.
  \ene 
   
We summarize the results that we have obtained in the following theorem.

\begin{theorem}\label{theo15.2} Suppose that $ V$ satisfies \eqref{1.6}, \eqref{1.7}. Then,  for $ E <0$, $ \xi(E;H_{A,B}, H_0)$ is piecewise constant,  it assumes  integral values, and \eqref{15.25}, \eqref{15.26} and \eqref{15.27} hold. For $ E >0$  
 \eqref{15.22} is valid, where $\Theta(k)$ is the continuous branch of $\textrm{arg}[\textrm{det}\, S(\sqrt{E})]$ that satisfies \eqref{15.21}.
 Moreover, \eqref{15.23} and \eqref{15.24} are true. Suppose that the function $f$ has two locally bounded derivatives and that for some $ \varepsilon >0$,
 \beq\label{15.28}
 f'(E)=O\left(\frac{1}{E^{1+\varepsilon}}\right),\, f''(E)= O\left(\frac{1}{E^{2+\varepsilon}}\right), \qquad \hbox{\rm as}\, \,E \rightarrow \infty.
 \ene
 Then $f(H_{A,B})- f(H_0)$ is trace class and
 \beq\label{15.29}
  \textrm{Tr}\left( f(H_{A,B})- f(H_0)\right) = \int_{-\infty}^\infty\, \xi(E; H_{A,B}, H_0)\, f'(E)\, dE.
 \ene
 \end{theorem} 
 
 \noindent{\it Proof:} it only remains to prove \eqref{15.29}. However, this formula follows from \eqref{15.7} and applying Theorem 9.7 in page 50 of \cite{ya2} and Remark 9.9 in page 51 of \cite{ya2} to    $f(H_{0,A,B})- f(H_0)$ and to$f(H_{A,B})- f(H_{0,A,B})$  using, respectively, \eqref{15.1} and \eqref{15.2} .

  If $V$ satisfies,
  \begin{equation}
\int_0^\infty
dx\, (1+x)\,   ||V(x)||<+\infty, \label{15.30}
\end{equation} 
 the total number of bound states, including multiplicity, $ \mathcal N, $ is finite (see Theorems ~ 8.1 and 8.6 of \cite{21}). Then, by \eqref{15.26}
  \beq \label{15.31}
  \xi(0_-; H_{A,B}, H_0)=: \lim_{E \uparrow 0}\,   \xi(E; H_{A,B}, H_0)= - \mathcal  N.
  \ene
   In the following therem we give a Levinson's theorem for the spectral shift function.
  \begin{theorem}\label{theo15.3}
  Suppose that $ V$ satisfies \eqref{1.7}, \eqref{15.30},
  and let $ \mathcal N < \infty$ be the number of bound states of $H_{A,B}$  including multiplicities. Then, denoting $ \xi(0+; H_{A,B}, H_0):= \lim_{E \downarrow 0}\,   \xi(E; H_{A,B}, H_0)$,
  \beq \label{15.27b}
  \xi(0+, H_{A,B}, H_0)= \frac{1}{2}\, [ n- \mu] -  \mathcal N,
  \ene
  where $ \mu$ is the (algebraic and geometric) multiplicity of the eigenvalue $+1$ of the zero-energy scattering matrix  $S_{A,B}(0)$. 
 \end{theorem}
 
 \noindent{\it Proof: }
Equation \eqref{15.27} follows from equation (9.13) of \cite{21}, \eqref{15.21} and \eqref{15.22}.

\end{document}